\documentclass[11pt,a4paper,oneside]{article}
\usepackage{jheppub}
\usepackage{graphicx,amssymb,amsmath,color}
\usepackage[normalem]{ulem}

\def\bea{\begin{eqnarray}}
\def\eea{\end{eqnarray}}

\def\bit{\begin{itemize}}
\def\eit{\end{itemize}}

\def\l{\left}
\def\r{\right}

\def\baa{\begin{array}}
\def\eaa{\end{array}}

\def\simgt{\mathrel{\lower2.5pt\vbox{\lineskip=0pt\baselineskip=0pt
           \hbox{$>$}\hbox{$\sim$}}}}
\def\simlt{\mathrel{\lower2.5pt\vbox{\lineskip=0pt\baselineskip=0pt
           \hbox{$<$}\hbox{$\sim$}}}}

\def\bfc{\begin{figure}\begin{center}}
\def\efc{\end{center}\end{figure}}
\def\nn{\nonumber\\}

\def\gsim{\gtrsim}
\def\lsim{\lesssim}

\def\GeV{{\rm GeV}}
\def\TeV{{\rm TeV}}
\def\eV{{\rm eV}}

\newcommand{\beq}{\begin{equation}}
\newcommand{\eeq}{\end{equation}}

\definecolor{chromeyellow}{rgb}{1.0, 0.65, 0.0}
\definecolor{darkcoral}{rgb}{0.8, 0.36, 0.27}
\definecolor{cadmiumgreen}{rgb}{0.0, 0.42, 0.24}
\definecolor{purple}{rgb}{0.7, 0, 0.7}

\title{\boldmath Bubble-assisted Leptogenesis}

\author[a]{Eung Jin Chun}
\author[a]{Tomasz P. Dutka}
\author[a,b]{Tae Hyun Jung}
\author[c]{Xander Nagels}
\author[c]{Miguel Vanvlasselaer}

\affiliation[a]{School of Physics, Korea Institute for Advanced Study, Seoul, 02455, Republic of Korea}
\affiliation[b]{Particle Theory  and Cosmology Group, Center for Theoretical Physics of the Universe,
Institute for Basic Science (IBS),
 Daejeon, 34126, Korea}
\affiliation[c]{Theoretische Natuurkunde and IIHE/ELEM, Vrije Universiteit Brussel, \& The  International Solvay Institutes, Pleinlaan 2, B-1050 Brussels, Belgium}

\emailAdd{ejchun@kias.re.kr}
\emailAdd{tdutka@kias.re.kr}
\emailAdd{thjung0720@gmail.com}
\emailAdd{Xander.Staf.A.Nagels@vub.be
}
\emailAdd{
miguel.vanvlasselaer@vub.be}

\abstract{We explore the possibility of embedding thermal leptogenesis within a first-order phase transition (FOPT) such that RHNs remain massless until a FOPT arises. Their sudden and violent mass gain allows the neutrinos to become thermally decoupled, and the lepton asymmetry generated from their decay can be, in principle, free from the strong wash-out processes that conventional leptogenesis scenarios suffer from, albeit at the cost of new washout channels. To quantify the effect of this enhancement, we consider a simple setup of a classically scale-invariant $B-L$ potential, which requires three RHNs with similar mass scales, in the ``strong-washout'' regime of thermal leptogenesis. Here we find that parameter space which requires $M_N\sim 10^{11}\text{ GeV}$ without bubble assistance is now predicted at $M_N \sim 5\times 10^9 \text{ GeV}$ suggesting a sizeable reduction from bubble effects. We numerically quantify to what extent such a framework can alleviate strong-washout effects and we find the lower bound on the RHN mass, $M_N \sim 10^{7}\text{ GeV}$, below which bubble-assisted leptogenesis cannot provide an enhancement. We also study the signature possibly observable at GW terrestrial interferometers and conclude that bubble-assisted leptogenesis models with relatively light masses, $M_N \lesssim 5\times 10^9 \text{ GeV}$ may be probable.}

\preprint{CTPU-PTC-23-17}

\arxivnumber{2305.10759}

\begin{document}
\maketitle
\flushbottom
\newpage

\section{Introduction}

Within the inflationary paradigm, the present universe is mostly independent of initial conditions. This forces us to consider a dynamical origin for the observed imbalance between matter and antimatter. Planck data and models of the early universe's evolution lead to a highly accurate prediction of the ratio~\cite{Ade:2015xua}
\bea
Y_B \equiv \frac{n_B -n_{\bar{B}}}{s}\bigg|_0 = (8.75 \pm 0.23)\times 10^{-11},
\eea 
where $n_B$ and $n_{\bar{B}}$ correspond to the number density of baryons and antibaryons respectively, $s$ corresponds to the entropy density, and the subscript denotes present time.

An elegant mechanism to generate the baryon asymmetry dynamically is through the decay of a heavy singlet fermion which carries lepton number, known as thermal leptogenesis~\cite{Fukugita:1986hr}. Here the baryon asymmetry arises from a dynamically generated lepton asymmetry via electroweak sphaleron processes, active within $T \in [10^{2}, 10^{12}]$ GeV. The source of the lepton asymmetry can be elegantly linked to the CP-violating decays of right-handed neutrinos (RHNs) in the type-I seesaw mechanism~\cite{Minkowski:1977sc,10.1143/PTP.64.1103,Mohapatra:1979ia,Glashow:1979nm,GellMann:1980vs}. The out-of-equilibrium condition, necessary for successful baryogenesis, can be naturally provided by the expansion of the universe\cite{doi:10.1146/annurev.ns.33.120183.003241,Riotto:1998bt}; as the temperature drops below the mass of the lightest RHN, the RHN decays remain efficient whereas the inverse process becomes Boltzmann suppressed (see for example \cite{Davidson:2008bu} for a review).

A convenient, but na\"ive, parameterization of the generated baryon asymmetry in conventional thermal leptogenesis is
\bea
Y_B = Y^{\rm eq}_{N}\, \epsilon_{\rm CP}\, \kappa_{\rm sph}\, \kappa_{\rm wash} ,
\label{eq:thermalleptorelation}
\eea
where $Y^{\rm eq}_{N} \equiv n^{\rm eq}_{N}/s$ is the relativistic abundance of the relevant RHN denoted by $N$, $\epsilon_{\rm CP}$ is the CP asymmetry in that RHN's decay, $\kappa_{\rm sph}$ accounts for the fraction of the lepton asymmetry which is converted to a baryon asymmetry by sphalerons (which has flavor-dependency), and $\kappa_{\rm wash} \leq 1$ accounts for active processes which washout the final asymmetry (typically dominated by the inverse-decay of $N$). We note however, that depending on the temperature regime of asymmetry creation and the specific coupling structure of the lightest RHN, Eq.~\eqref{eq:thermalleptorelation} should be modified with flavour effects to properly estimate the asymmetry. However, it allows for a good qualitative description of the physics parameters which affect the final asymmetry generated.

An estimate for the typical scale of thermal leptogenesis, ignoring flavor effects for qualitative simplicity, can be derived as follows. The CP asymmetry parameter is bounded from above~\cite{Davidson:2002qv} for a hierarchical spectrum of RHNs: $\left|\epsilon_{\rm CP}\right| \lesssim (3 M_N (m_{\nu_3}-m_{\nu_1}))/(8\pi v_h^2)$ where $v_h=246$ GeV is the Standard Model Higgs vacuum expectation value (vev). Taking $\kappa_{\rm sph} = 28/79$, Eq.~\eqref{eq:thermalleptorelation} can be rearranged to
\bea
M_N \sim \frac{10^{9}\text{ GeV}}{\kappa_{\rm wash}}
\l( \frac{0.05\,\eV}{m_{\nu_3}-m_{\nu_1}}
\r),
\eea
where $\kappa_{\rm wash}$ cannot be analytically determined.
The washout factor depends on the parameter  $K \equiv \Gamma_D/H(M_N)\simeq m_\nu/10^{-3} \text{ eV}$ where $m_\nu$ denotes the effective neutrino mass scale related to the couplings of the RHNs to the standard model neutrinos.  A weak-washout scenario corresponds to $K \lesssim 1$, where $\kappa_{\rm wash} \simeq 1$, in which case $M_N$ is bounded from below at roughly $10^9\text{ GeV}$, famously known as the Davidson-Ibarra bound~\cite{Davidson:2002qv}.  When $m_\nu$ is taken to be the atmospheric (solar) neutrino mass scale, $\simeq 0.05\, (0.01)$ eV which occurs for example with democratic couplings of $N$ to $\nu$, we have $K\approx 50\, (10)$ corresponding to a ``strong washout regime'' of thermal leptogenesis. Detailed numerical calculations imply $\kappa_{\rm wash} \sim 10^{-2}$--$10^{-3}$~\cite{Buchmuller:2004nz}. From this, we obtain the rough scale, $M_N \sim 10^{11}\text{ GeV}$ for successful strong-washout leptogenesis. Accounting for flavour effects in the thermal bath can change this rough estimation for some choices of parameters~\cite{Abada:2006fw,Abada:2006ea,DiBari:2018fvo} with a special role played by the possible Majorana phases\,\cite{Blanchet:2006be,
Blanchet:2008pw}. This is important in some models of leptogenesis, for example for $N_2$ leptogenesis scenarios, where detailed flavored calculations are necessary, but usually weakens the washout effect by an order-one factor. However, the majority of the parameter space of type-I seesaw thermal leptogenesis requires RHN masses much larger than what is implied by the Davidson-Ibarra bound, as can be seen in Appendix B of~\cite{Moffat:2018wke} or the numerical results of~\cite{Granelli:2020ysj}.

%\medskip

As a consequence of the high energy scales required, testing the minimal (hierarchical) model is challenging to say the least. On the other hand, if $B-L$ is promoted to a good symmetry which is broken spontaneously and this transition is first-order, in principle we may be able to indirectly test for leptogenesis by searching for gravitational wave signals produced during bubble percolation.
However, most of the parameter space (strong-washout leptogenesis) remains outside the sensitivity range of future gravitational wave detectors as
the peak frequency is expected to be high, $f_{\rm peak} \sim 10^5$--$10^6\,{\rm Hz}\, (T_*/10^{11}\,\GeV)$, 
where $T_*$ is the reheating temperature of the phase transition. 
Without modifications, one of which we explore in this work, the prospects for testability are bleak.

We consider bubble dynamics during a first-order phase transition (FOPT) as a source of a strong departure from thermal equilibrium on the RHN population.
We show that, in this scenario, the required value of $M_N$ is more than one order of magnitude lower than that of the conventional scenario when requiring successful leptogenesis, and therefore it is within the testable range of future gravitational wave detectors.
The idea of utilizing bubble dynamics for baryogenesis through a sudden mass gain, to be described below, was first proposed in~\cite{Baldes:2021vyz}\footnote{However in \cite{Baldes:2021vyz}, the mass gain mechanism was applied to a model using the decay of coloured scalars.} and was applied to leptogenesis recently in~\cite{Huang:2022vkf,Dasgupta:2022isg}\,\footnote{
There are various other implications of the strong departure from thermal equilibrium induced by FOPTs. For example, \cite{Cline:2022xhx} for the case of a supersymmetric phase transition, ~\cite{Azatov:2021irb, Baldes:2021vyz, Azatov:2022tii} for baryogenesis scenarios without leptogenesis and~\cite{Hui:1998dc,Chung:2011hv, Chung:2011it, Hambye:2018qjv,Falkowski:2012fb,Baker:2019ndr, Chway:2019kft, Azatov:2021ifm,Wong:2023qon,
Elor:2021swj, Baldes:2022oev, Guo:2015lxa} in the context of dark matter.
}. We will provide a complementary analysis by performing a numerically detailed scan of such a leptogenesis scenario, specifically for the non-resonant case, to evaluate the final baryon asymmetry  as well as potential gravitational wave (GW) signatures.
% {\color{purple}
% In conclusion, we find that the typical mass scale of RHNs required for the observed baryon asymmetry is a few times $10^9\,\GeV$, and GW signals produced by the FOPT are within the detectable range of future GW detectors, which is opposed to the conclusion made in Ref.\,\cite{Huang:2022vkf}.
% }{\color{red} feels weird to have this in the introduction}

The setup of the scenario we consider is as follows: we assume that the Majorana mass of the RHNs are provided by the vev of a scalar field and the phase transition corresponding to this spontaneous symmetry breaking is first-order (see Ref.\,\cite{Shuve:2017jgj} for the study of the second-order case).
The relevant part of the Lagrangian can be written in the mass basis of the RHNs as
\begin{align}
\label{Eq:L_RHN}
\mathcal{L}_{\text{int}} = &\frac{1}{2} \sum_{I} 
y_I \Phi \bar{N}_I^c N_I + \sum_{\alpha,\, I} Y_{D,\alpha I} H\bar{L}_{\alpha}N_I +  h.c.,
\end{align}
where $L_\alpha$ are the SM lepton doublets, $N_I$ are the three 
families of heavy right-handed neutrinos, $Y_{D,\alpha I}$ are the \emph{Dirac} Yukawa couplings between $N_I$ and $L_\alpha$, and $y_I$ are \emph{Majorana} Yukawa couplings. After the phase transition, $\langle \Phi \rangle \equiv v_\phi/\sqrt{2}$, and the type-I seesaw Lagrangian is recovered with $M_I = \frac{1}{\sqrt{2}}y_I v_\phi$.
As we assume that the critical temperature of the $\Phi$ phase transition is much greater than that of the electroweak critical temperature, we assume $\langle H \rangle = 0$ during and after the FOPT but fix $\langle H \rangle \neq 0$ where appropriate.

\begin{figure}
      \centering
      \includegraphics[width=0.95\textwidth]{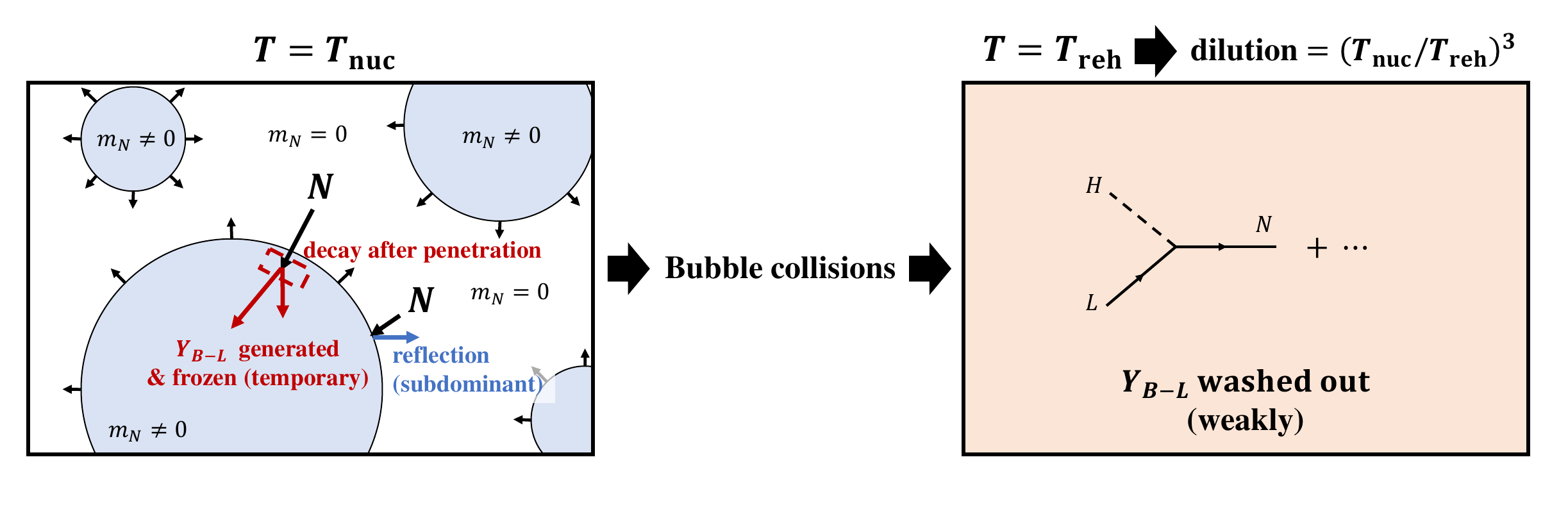}
      \caption{ Schematic picture of the bubble-assisted leptogenesis scenario during bubble expansion (left) and after bubble collisions (right). 
      }
      \label{fig:schematic}
\end{figure}

With these assumptions, the typical temperature evolution of the conventional thermal leptogenesis scenario can be modified in the following way
(see Fig.\,\ref{fig:schematic} for the summary of our process): RHNs are massless (ignoring thermal effects) until the phase transition of $\Phi$ after which they suddenly become massive within the bubbles of the broken phase. As soon as $M_I \neq 0$, the RHNs decay and generate a nonzero lepton asymmetry very rapidly, owing to the strong-washout regime typically predicted in the type-I seesaw. If the bubble nucleation temperature $T_{\rm nuc}$ is significantly smaller than the masses of the RHNs, the inverse decays within the bubbles will be immediately Boltzmann suppressed and, in principle, $\kappa_{\rm wash} \sim \mathcal{O}(1)$. We call this scenario ``bubble-assisted leptogenesis''. A strong assumption we require in our example setup of bubble-assisted dynamics is for all three RHNs to have comparable masses which we explain in Section~\ref{sec:formalism}, however this may not be a generic requirement and we stress that we only require a confluence of scales for the three masses. Similarly, a large degeneracy in their masses, \textit{\`a la} resonant leptogenesis, is not required nor problematic for the setup in any way but will simply introduce an independent source of asymmetry enhancement.
The enhancement of $Y_B$ we quantify is specifically that which is a result of the out-of-equilibrium effects catalyzed by bubble dynamics, and the results we present throughout the paper will not depend on the choice of $\epsilon_{\rm CP}$. 

Unsurprisingly, the scenario introduces new dynamics which can severely affect this simple qualitative picture and must be properly estimated. During the bubble expansion, the latent heat stored in the scalar potential $\Delta V$, which is the difference in the scalar potential energy density between the true and the false vacuum,  is converted to a combination of bubble wall kinetic energy and fluid bulk motion. After the bubble walls and fluid shells collide, this latent heat reheats the background plasma, increasing the temperature to $T_{\rm reh}\sim (\Delta V+\rho_{\rm plasma}(T_{\rm nuc}))^{1/4}$. 

The reheating affects our scenario in two ways:
i) as the asymmetry is generated during the bubble expansions, it will be diluted by a factor of $(T_{\rm nuc}/T_{\rm reh})^3$ due to reheating, ii) in order to avoid a strong washout, we also require $M_N/T_{\rm reh} \gg 1$ along with $M_N/T_{\rm nuc} \gg 1 $, otherwise the RHN inverse decays become rapid after the bubbles collide.

In general, there is a close correlation between $T_{\rm nuc}/T_{\rm reh}$ and $M_N/T_{\rm reh}$ for a given scalar sector. Since $T_{\rm reh} \sim (\Delta V+\rho_{\rm plasma}(T_{\rm nuc}))^{1/4}$ and $M_N = \frac{1}{\sqrt{2}} y v_\phi$, requiring a large $M_N/T_{\rm reh}$ means that the potential should be quite flat, i.e. $\Delta V/v_\phi^4 \ll 1$. This kind of flat potential, in general, results in a strong supercooling, so there is always a tension between the dilution factor vs the washout factor for a given scalar sector.

Another complication of the scenario comes from new processes which can lead to additional suppression such as the unavoidable annihilation process $N N \to \phi \phi$ with $\phi$ the radial mode of $\Phi$\,\footnote{
In~\cite{Huang:2022vkf} the RHN mass was restricted to be large enough to totally avoid these annihilation processes, but we find that their criterion was too conservative. On the other hand, Ref.~\cite{Dasgupta:2022isg} studied $\mathcal{O}(10)\,\TeV$ (and more recently \cite{Borah:2023saq} for intermediate scale leptogenesis) scale right-handed neutrino masses where the annihilation rate is typically much more dominant than the decay rate.
In this case, there can be a sizable suppression in the final baryon asymmetry compared to the estimation presented in Ref.\,\cite{Dasgupta:2022isg}. 
}.
Due to these new annihilation channels, which are absent in conventional thermal leptogenesis, these extra channels further deplete the population of RHNs, i.e. some will annihilate instead of decaying. To distinguish this new effect from the conventional washout process in thermal leptogenesis, we will refer to these new channels as `depletions'. We quantify the impact that these depleting interactions have on the final asymmetry through the parameter $\kappa_{\rm dep}$.

In summary, we are utilizing bubble dynamics as a means to increase $\kappa_{\rm wash}$, which appears in conventional thermal leptogenesis, but this necessarily introduces a new suppression factor $\kappa_{\rm dep} (T_{\rm nuc}/T_{\rm reh})^3$. Considering all these effects, we quantitatively study the amount of enhancement that can be achieved in the scenario as described above.

A qualitative summary of the different steps necessary to evaluate the final asymmetry, to be discussed in detail in Section~\ref{sec:formalism}, is as follows:
\begin{itemize}
    \item[1.]{\bf Estimation of $N_I$'s penetration rate into the bubbles.}
    \\
    We first obtain the velocity of the bubble walls, for a given choice of parameters in the scalar potential, 
    and integrate the distribution function of RHNs which have enough momentum to enter the bubbles.
    The amount of RHNs which enter the bubbles will be reduced by a factor of $\kappa_{\rm pen}$\,\footnote{
    We find that, in most of the parameter space, 
    the bubble wall runs away and we can avoid a large suppression coming from the reflection of $N$.
    This opposes to the argument made in Ref.\,\cite{Shuve:2017jgj}.}.
    
    \item[2.]{\bf Evolution of $Y_{B-L}$ before bubble collisions}
    \\
    We solve the Boltzmann equations for $Y_{B-L}$ and $Y_{N_I}$ inside the bubbles 
    including the usual washout of $N_I$ as well as the depleting channels which now occur, e.g. $N_IN_I \to \phi \phi, f\bar f$. 
    The onset of leptogenesis is now $T=T_{\rm nuc}$, and the time of this process is limited 
    by the duration of the PT: $\Delta t_{\rm PT} \sim \mathcal{O}(10^{-2}) H^{-1}$, where $H$ is the Hubble rate. The suppression in this step can be mostly encapsulated by $\kappa_{\rm dep}$, however there will also be a less-dominant contribution from $\kappa_{\rm wash}$. 
    
    \item[3.]{\bf Evolution of $Y_{B-L}$ after bubble collisions.}
    \\
    We solve the Boltzmann equation again, now starting from $T=T_{\rm reh}$ with boundary conditions 
    obtained from the previous step with a dilution factor of $(T_{\rm nuc}/T_{\rm reh})^3$.
    We estimate $\kappa_{\rm wash}$ as the suppression factor coming from this evolution as $\kappa_{\rm dep}$ will not contribute due to the negligible population of RHNs.
    We take the asymptotic value of $Y_{B-L}$ and multiply it by the sphaleron conversion factor 
    to obtain the final $Y_B$.
\end{itemize}

Combining the result of each step, the final baryon asymmetry in the bubble-assisted leptogenesis mechanism can be expressed as
\bea
Y_B = Y^{\rm eq}_N
\epsilon_{\rm CP} \,
\kappa_{\rm sph} \,
\kappa_{\rm pen} \,
\kappa_{\rm dep} \,
\kappa_{\rm wash} \,
\left( \frac{T_{\rm nuc}}{T_{\rm reh}}
\right)^3,
\label{eq:finalasymparam}
\eea
where $Y^{\rm eq}_N$ is population of RHNs outside the bubbles.
For our numerical calculation, we solve the full Boltzmann equations from which each factor can be inferred.

Within this framework, in order to provide a concrete numerical example, we consider the classically scale-invariant setup for the scalar sector\,\cite{Buchmuller:2013lra, Chun:2013soa, Jinno:2016knw, azatov2020phase, Marzo:2018nov, Bian:2019szo}, where the symmetry breaking is induced by either an additional real singlet scalar field or from a minimal gauged $U(1)_{B-L}$ (see Section\,\ref{sec:results} for details).

\begin{figure}[t]
      \centering
      \includegraphics[width=0.45\textwidth]{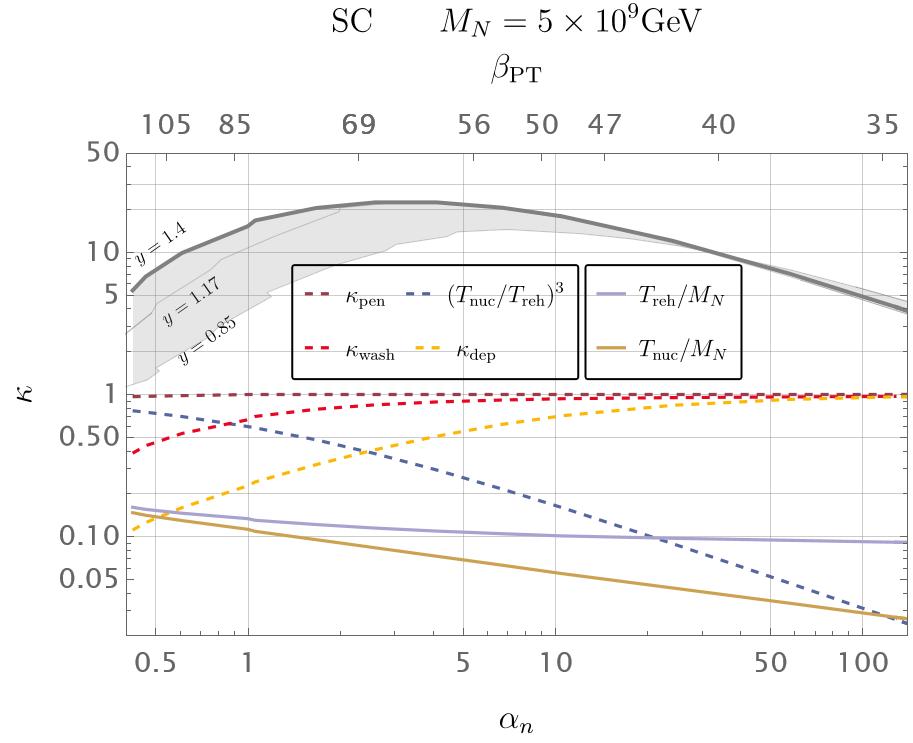}
      \includegraphics[width=0.45\textwidth]{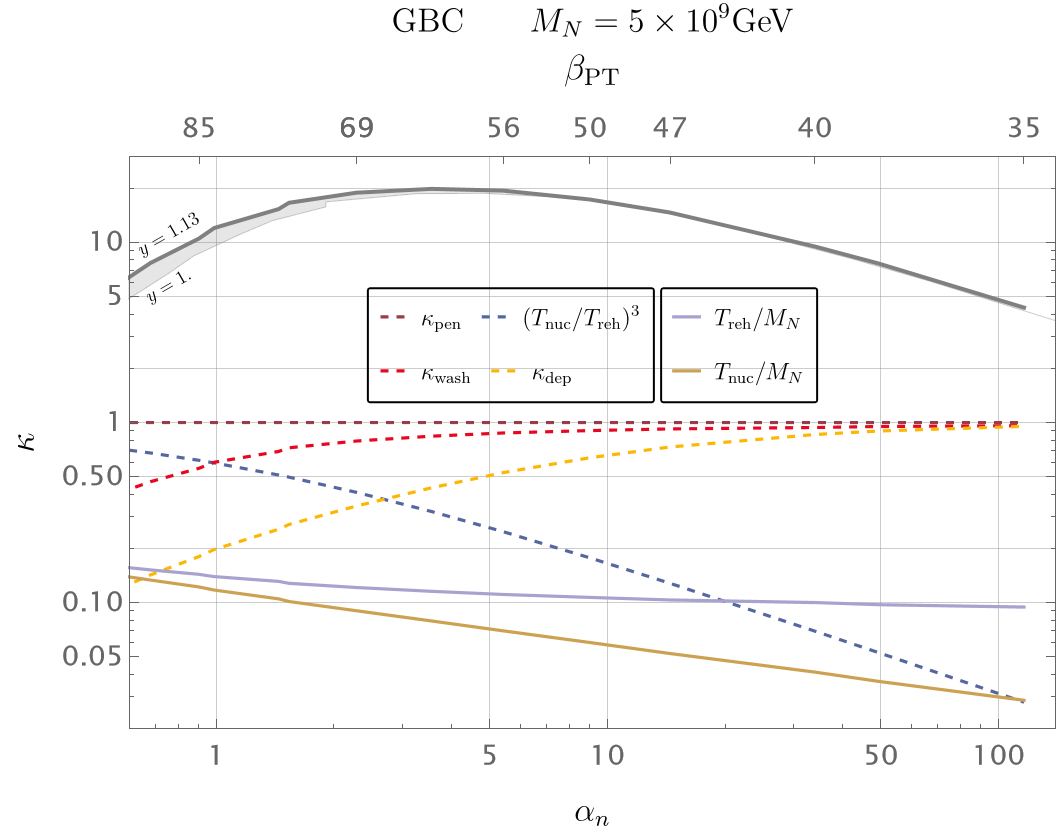}
      \caption{Intuitive presentation of the enhancement for the scalar catalyzed model (left panel) and the gauge boson catalyzed model (right panel) as a function of the PT strength $\propto \alpha_n$. 
      The enhancement of $Y_B$ in the bubble-assisted leptogenesis scenario compared to the conventional thermal leptogenesis scenario is depicted by the grey band for different values of $y$ whilst keeping the overall RHN mass fixed. 
      We also depict how the suppression factors within the bubble-assisted leptogenesis scenario: $\kappa_{\rm pen}$, $\kappa_{\rm wash}$, $\kappa_{\rm dep}$ and $(T_{\rm nuc}/T_{\rm reh})^3$ vary with the strength, Eq.~\eqref{eq:alpha}. 
      For concreteness, we fixed $y$ to the largest value displayed on the figure, corresponding to largest enhancement, when plotting these suppression factors.
      The timescale, Eq.~\eqref{eq:beta_PT}, of the phase transition is also displayed. At small values of $\alpha_n$, for weaker phase transitions, a larger fraction of $N$ are reflected against the bubble ($\kappa_{\rm pen}$ begins to decrease) and the transition is less strongly out-of-equilibrium. 
      This implies that $M_N/T_{\rm reh} \leq 1$ so suppression from wash-out and depletion significantly reduces the final asymmetry.  
      At larger values of $\alpha_n$, corresponding to stronger phase transitions, supercooling begins which leads to a strong dilution of the final asymmetry as $T_{\rm nuc} \ll T_{\rm reh}$. 
      In between those two suppressive regimes, we observe a peak in the enhancement for $\mathcal{O}(1)$ values of $\alpha_n$.}
      \label{fig:summary}
\end{figure}

Fig.\,\ref{fig:summary} shows a short summary of our results where the left and right panels correspond to the scalar catalyzed case (SC) and the gauge boson catalyzed case (GBC), respectively. Here, the grey bands show the amount of enhancement we obtain compared to the conventional thermal leptogenesis, and the horizontal axis shows the strength of the supercooling, $\alpha_n$ which is defined in Eq.~\eqref{eq:alpha}.
We obtain a $\mathcal{O}(20)$ enhancement compared to conventional scenarios for $M_N = 5\times 10^9\,\GeV$.

We can understand this enhancement in terms of $\kappa_{\rm pen}$, $\kappa_{\rm wash}$, $(T_{\rm nuc}/T_{\rm reh})^3$, and $\kappa_{\rm dep}$ which are depicted by dashed curves. 
The amount of RHNs penetrating into the bubble, $\kappa_{\rm pen}$, mostly stays order one in this parameter space, but slightly decreases when $\alpha_n \lsim 1$. For $\alpha_n >5$, $\kappa_{\rm wash}\simeq 1$, so the standard washout suppression inherent in thermal leptogenesis is circumvented. 
We find that $\kappa_{\rm dep}$ causes a stronger suppression compared to $\kappa_{\rm wash}$, highlighting the importance of including these new annihilation channels. Both $\kappa_{\rm dep}$ and $\kappa_{\rm wash}$ sharply decrease as $\alpha_n$ decreases. In the gauged case, there is an additional contribution to the annihilation $\kappa_{\rm dep}$ coming from the $s$-channel process $N_I N_I \to q \bar q, \, l \bar l$. 

At large values of $\alpha_n$, the washout and depletion effects may be small, but the dilution factor from reheating, $(T_{\rm nuc}/T_{\rm reh})^3 = (1+\alpha_n)^{-3/4}$, strongly suppresses the final asymmetry. 
For large enough values of $\alpha_n$, the bubble-assisted scenario will in fact predict a suppressed final asymmetry compared to the typical thermal scenario. 
The final asymmetry is proportional to the all these factors and we find the enhancement is maximized around $\alpha_n \sim 5$.

We also investigate the possibility of testing this scenario via gravitational wave detectors, as the collisions of the bubbles required by this scenario necessarily generate gravitational waves. This has also been studied in the same context, but in different parameter regions, in Refs.~\cite{Huang:2022vkf,Dasgupta:2022isg,Borah:2023saq}.
As we find that bubble-assisted leptogenesis is viable for low values of $M_N$ (and therefore low values of the symmetry-breaking scale), the peak frequency of gravitational waves produced can be within the observable range of terrestrial observers like ET\cite{Maggiore:2019uih}, CE\cite{Evans:2021gyd} and LIGO O5 \cite{LIGOScientific:2007fwp}.
Larger values of $M_N$, which are still viable in the bubble-assisted leptogenesis scenario, would require detectors that can probe higher frequency ranges, such as those proposed in Refs.~\cite{Aggarwal:2020umq,Domcke:2022rgu,Chou_2017,Goryachev:2014yra,Berlin:2021txa,Berlin:2023grv} albeit with improved sensitivities.

This paper is organized as follows.
Section\,\ref{sec:review} briefly reviews how we treat the FOPT and introduces several effective parameters that are relevant to our leptogenesis scenario.
Section\,\ref{sec:formalism} provides the framework to estimate the net baryon asymmetry in a systematic way based on these effective parameters.
Section\,\ref{sec:results} shows numerical results for the case of the classically scale-invariant scalar sector.
We discuss gravitational wave signals in Section\,\ref{sec:GW}, and conclude in Section\,\ref{sec:conclusion}.

\section{Brief review on cosmological first-order phase transition}
\label{sec:review}
We remain agnostic to the tree-level potential of $\Phi$ until we perform the numerical scans in Section\,\ref{sec:results} as our formalism, particularly Section\,\ref{sec:formalism}, applies to an arbitrary scalar sector of $\Phi$ provided that $\Phi \bar{N}^c N$ exists. Here we summarize the general formalism used to extract the effective parameters of a FOPT for a given scalar potential.

\subsection{Temperature-dependent effective potential}
\label{sec:FOPT}
For a given tree-level Lagrangian of $\Phi$, 
the effective potential acquires quantum corrections at zero temperature.
The one-loop contribution from a particle $i$  can be written as\,\cite{PhysRevD.7.1888}
\begin{equation}
 	V_{CW}(m_i^2(\phi)) = 
  (-1)^{2s_i} g_i\frac{m_i^4(\phi)}{64\pi^2}\Big[\log \Big(\frac{m_i^2(\phi)}{\mu^2}\Big)-c_i\Big],
\label{eq:CW}
\end{equation}
where $\phi$ is the real part of $\Phi$,  $\phi \equiv \sqrt{2} {\rm Re}(\Phi)$,
$m_i(\phi)$ and $s_i$ are the $\phi$-dependent mass and spin of a particle $i$ with $g_i$ degrees of freedom.
Here $\mu$ is the renormalization scale, and $c_i$ is a constant depending on the subtraction scheme.
In this work we use the $\overline{\rm MS}$ scheme where $c_i=3/2$ for $s_i \in \{0,1/2\}$ and $c_i=5/6$ for $s_i=1$.

At a finite temperature $T$, thermal effects can be captured by including the \emph{thermal potential} of the form 
\begin{equation}
 	V_T(m_i^2(\phi)) = \pm \frac{g_i}{2\pi^2}T^4 J_{\rm B, F} \Big(\frac{m_i^2(\phi)}{T^2}\Big) \quad\mbox{with}\quad
  J_{\rm B, F}(y^2) = \int \limits_0^{\infty} dx \,\, x^2\log \Big[1 \mp \exp{(-\sqrt{x^2+y^2})}\Big],
\end{equation}
where the upper (lower) sign is for the bosonic (fermionic) case.
In our numerical calculations we use the full form for $J_{\rm B,F}$. For illustrative purposes, we show the expansions of $J_B(y^2)$ \cite{Curtin:2016urg}:
 	\begin{equation}
 	J_B(y^2 \ll 1) \approx -\frac{\pi^2}{45}+\frac{\pi^2}{12}y^2-\frac{\pi}{6}y^3+ \cdots ,
        \qquad J_B(y^2\gg 1) \approx  - \sum_{n = 1}^{20}\frac{1}{n^2}y^2K_2(y\cdot n) ,
 	\end{equation}
where $K_2(z)$ is the Bessel function of the second-kind. 
Note that bosonic interactions are required to make the phase transition first order due to the $y^3$ term in  $J_B(y^2\ll 1)$.  

We also incorporate the resummation of  the self-energy diagrams of bosons following the so-called truncated-full-dressing procedure\,\cite{Curtin:2016urg}.
This can be effectively done with the replacement
\bea
m_i^2(\phi) \to m_i^2(\phi)+\Pi_i,
\eea
where $\Pi_i$ is the thermal correction to the mass of a particle $i$ (see also Ref.~\cite{Schicho:2022wty} for an updated tool for the resummation). Theoretical uncertainties related to the use of the perturbative method have been discussed in Ref.~\cite{Croon:2020cgk} for polynomial potentials, however, we expect them to be mild in the scalar potential that we use in Section.~\ref{sec:model}. 
We find that there are no significant changes to our results with this replacement, so our parameter space is numerically stable.

The complete, temperature-dependent, scalar potential we consider is given by the sum of all these contributions:
\bea 
  V(\phi, T) = V_0(\phi) +\sum_i V_{CW} (m_i^2(\phi) +\Pi_i ) + \sum_i V_T(m_i^2(\phi)+\Pi_i),
\label{eq:VT}
\eea
where $V_0(\phi)$ corresponds to the tree-level potential.

\subsection{Bubble nucleation}

\begin{sloppypar}
Once the temperature-dependent scalar potential is calculated, we obtain the bounce solution and the bounce action using \texttt{Mathematica} based on the well-known overshoot/undershoot method. Additionally, we crosscheck our results against \texttt{CosmoTransition}~\cite{Wainwright:2011kj} and \texttt{FindBounce}~\cite{Guada:2020xnz}.
\end{sloppypar} 

The bubble nucleation rate can be estimated as the probability to have a critical bubble per unit time and unit volume, where the origin of this configuration can be either due to quantum or thermal fluctuations.
We approximate it as
\bea
\Gamma(T)\sim \text{max} \l[T^4 \l(\frac{S_3}{2\pi T}\r)^{3/2}\text{Exp}(-S_3/T),~~ R_0^{-4} \l(\frac{S_4}{2\pi}\r)^2 \text{Exp}(-S_4)\r],
\eea
where $S_3$ and $S_4$ are $O(3)$ (thermal) and $O(4)$ (quantum) bounce actions, respectively, and $R_0$ is the initial bubble radius. 
Since the bubble-assisted leptogenesis framework requires the FOPT not to be too strongly supercooled, the $O(3)$ bounce solution always dominates so we ignore contributions from $S_4$.

Now, let us discuss the timeline of the first-order phase transition.
At the critical temperature $T=T_{\rm crit}$, the two local minima are degenerate, and $\Gamma(T_{\rm crit})=0$.
So in any first-order phase transition there is a period where $\Gamma(T)$ is much slower than the Hubble expansion rate.
During this period, although bubbles can be nucleated, the expansion of spacetime is more efficient and the phase transition does not proceed.
The universe is supercooled until $\Gamma(T)$ becomes comparable to the Hubble rate, $H(T)^4$.

When $\Gamma(T)\sim H(T)^4$, the average distance of two nearby bubble nucleations becomes less than the horizon size and the bubble expansion can physically reduce the volume of the false vacuum. We define this temperature as $T_{\rm nuc}$, $\Gamma(T_{\rm nuc})\equiv H(T_{\rm nuc})^4$, which can be approximately calculated by solving
\bea
\frac{S_3(T_{\rm nuc})}{T_{\rm nuc}} 
\approx 4\log\left[\frac{T_{\rm nuc}}{H(T_{\rm nuc})}\right] ,
\label{estimateTn}
\eea 
where $H(T_{\rm nuc})$ should include the contribution of the vacuum energy that comes from the scalar potential.

Around this temperature, new bubbles are nucleated and expand.
The time scale of this procedure (roughly the time scale between the bubble nucleation temperature and the end of the phase transition) can be parameterized by
\bea
(\Delta t)^{-1}_{\rm PT} 
\sim \left. -\frac{d(S_3/T)}{dt} \right|_{T=T^{\rm nuc}}
\equiv H_{\rm reh} \beta_{\rm PT} ,
\label{eq:beta_PT}
\eea
where $H_{\rm reh}$ is the Hubble rate at the reheating temperature.
Note that $\beta_{\rm PT}$ is a dimensionless quantity unlike the conventional definition that can be found in most references.

For the classically scale-invariant potential we consider in Section~\ref{sec:results}, we numerically find that $\beta_{\rm PT} \sim 50$.
Since $\beta_{\rm PT}$ is typically large, the duration of the phase transition is much shorter than the Hubble time scale.
So, we can ignore the redshift of temperature during the bubble expansion and treat the temperature to be approximately constant, $T=T_{\rm nuc}$.

The strength of a FOPT is parameterized by
\bea
\alpha_n \equiv \frac{\Delta V}{\rho(T_{\rm nuc})},
\label{eq:alpha}
\eea
where $\rho(T)=\frac{\pi^2}{30}g_* T^4$ is the plasma energy density which has $g_*$ effective relativistic degree of freedom at $T$.
Since $\Delta V$ will eventually be converted to plasma energy after bubble collisions, we can obtain the reheating temperature by
\bea
\label{eq:reheatrelation}
\rho(T_{\rm reh}) \simeq \rho(T_{\rm nuc}) + \Delta V,
\eea
where we assumed the time scale of the reheating procedure is much shorter than the Hubble time scale\,\footnote{
If the reheating procedure instead occurs for longer than the Hubble time scale, $T_{\rm reh}$ is suppressed compared to Eq.~\ref{eq:reheatrelation}, which implies less dilution of the final baryon asymmetry.
}.
Since $\rho\propto T^4$, the dilution factor after bubble collisions is simply given by
\bea
\left( \frac{T_{\rm nuc}}{T_{\rm reh}}  \right)^3
\simeq (1+\alpha_n)^{-3/4}.
\eea

\section{Bubble-assisted leptogenesis}
\label{sec:formalism}
In the following, we assume that the masses of $N_I$ are nearly degenerate:
\bea
y\equiv y_1\simeq y_2\simeq y_3 \,\, \implies \,\,
M_N \equiv M_1\simeq M_2\simeq M_3.
\eea
If there is a hierarchical mass spectrum of $N_I$, $y_1 \ll y_3$, the mass of $N_1$ will generally be much smaller than $T_{\rm reh}$, at the very least for the classically scale-invariant potential we will assume in Section~\ref{sec:results}. For a hierarchical spectrum, the asymmetry generated through bubble dynamics will therefore be washed out by the inverse decays of $N_1$ after reheating, so the scenario approaches the predictions of conventional thermal leptogenesis (with additional depleting effects from $\phi$ interactions) as the hierarchy amongst $N_I$ increases. This requirement may be lifted if a different scalar potential is assumed so it may not be a necessary prediction of bubble-assisted dynamics. We also only consider parameter space where the three RHN decay widths are in the strong-washout regime, which is true for a very large fraction of the total parameter space~\cite{Moffat:2018wke}.

%We do not assume a tuning in the flavor structure of the Dirac Yukawa, $Y_D$, so the decay width for each $N_I$ is within the strong-washout regime of thermal leptogenesis and is similar in size.

\subsection{Penetration rate}
\label{subsec:penrate}

How efficiently the massless RHNs outside of the expanding bubbles can penetrate the bubble wall is an important ingredient, as their penetration causes a large departure from the equilibrium number density of (the now massive) RHNs within the bubbles.
An order-one penetration rate, $\kappa_{\rm pen}$, is desired for bubble-assisted leptogenesis.

The penetration rate is closely related to the bubble wall velocity. The fraction of heavy particles entering the wall will grow with the boost factor of the wall, $\gamma_w$, and reach an order one fraction when $M_N \lesssim \gamma_w T_{\rm nuc}$. 
The pressure also increases with the boost factor, requiring a stronger release of energy $\Delta V$, and thus favoring a larger $\alpha_n$.
Those considerations may result in a (mild) tension between $\kappa_{\rm pen}$ and the dilution factor $\propto \alpha_n^{3/4}$.

\begin{figure}[t]
      \centering
      \includegraphics[width=0.95\textwidth]{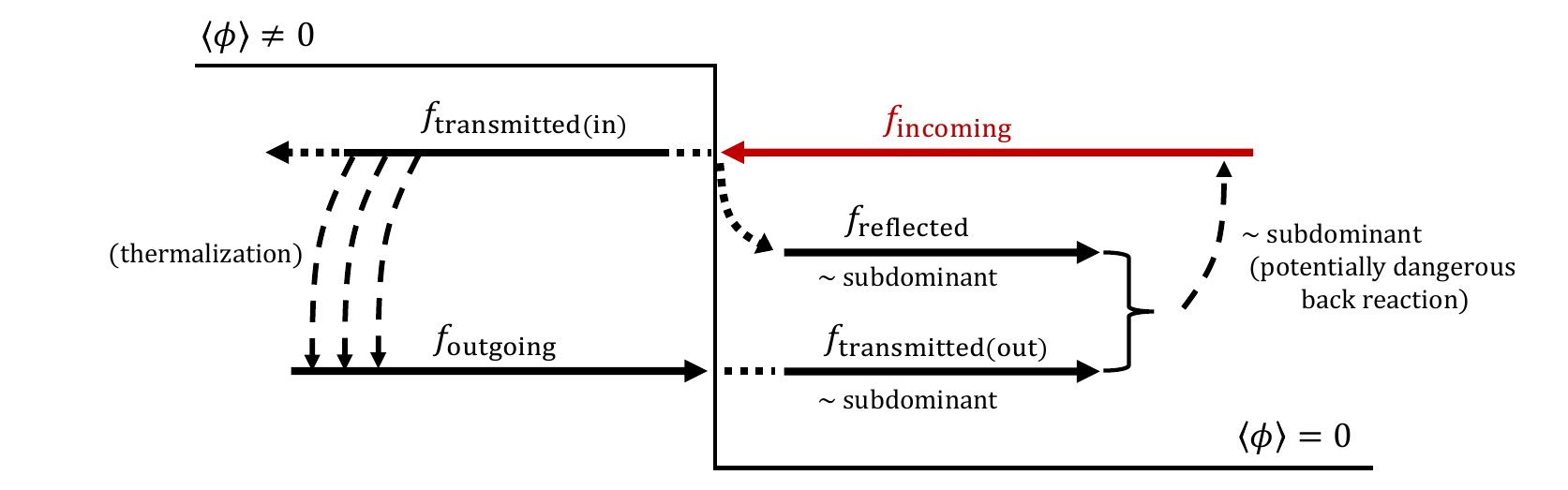}
      \caption{Schematic picture of particle distribution in the bubble wall rest frame. The arrows indicate the momentum direction of $f$'s.
      }
      \label{fig:f_incoming}
\end{figure}

To estimate $\kappa_{\rm pen}$ in a consistent way, we take the collisionless limit \cite{Dine:1992wr,Arnold:1993wc} which was recently reviewed in \cite{Mancha:2020fzw, Vanvlasselaer:2020niz} and is valid for fast walls $\gamma_w \gg 1$.
In the bubble wall rest frame, let us decompose the distribution of $N$ as Fig.\,\ref{fig:f_incoming} depending on the momentum directions.
We assume that particles are thermalized far out of the bubble with the distribution $f_{\rm incoming}$.
When incoming particles reach the bubble wall, they can be either reflected ($\to f_{\rm reflected}$) or transmitted ($\to f_{\rm transmitted(in)}$) depending on whether the longitudinal momentum is greater than the mass inside the bubble or not.
The transmitted particles get thermalized deep inside the bubble, and some of them change their momentum direction to escape the bubble with a distribution $f_{\rm outgoing}$.
Denoting the distribution of particles that have escaped $f_{\rm transmitted (out)}$, the total distribution outside the bubble with momentum direction aligned to the bubble wall expansion is given by $f_{\rm reflected} + f_{\rm transmitted(out)}$.
In the following calculation, we neglect the back-reaction of $f_{\rm reflected} + f_{\rm transmitted(out)}$ to the incoming distribution.
This assumption is self-consistent if the reflection rate is small.  We indeed focus on the parameter space where the bubble wall velocity is mostly relativistic (so $f_{\rm outgoing}$ is negligible) and $\kappa_{\rm pen} \simeq 1$ (so $f_{\rm reflected}$ is negligible).

In this limit, the thermal distribution of the incoming fluid (outside the bubbles) in the wall rest frame is given by 
\bea
f_{\rm incoming}\simeq\frac{1}{e^{\gamma (E + v p_z)/T_{\rm nuc}}\pm 1},
\eea
where $\pm$ signs are for Fermi-Dirac and Bose-Einstein distributions respectively, $v$ ($\gamma$) is the positively defined bubble wall velocity (boost factor), and $+z$ is the direction of the bubble wall expansion.

The pressure can be obtained by summing up the momentum transfer during reflections and transmissions of a particle from the outside and the inside.
Each contribution can be written in the form of 
\bea
\mathcal{P} = \int \frac{d^3p}{(2\pi)^3} \,\,
 (\Delta p) f
= \int \frac{ dp_z\, dp_\perp \,  2\pi p_\perp}{(2\pi)^3} \,\,
 (\Delta p) f,
\label{Pres:general}
\eea
where the momentum transfer $\Delta p$ and the integral range depend on whether particles are reflected or transmitted and from which side of the bubble wall they come from.
With these assumptions, using $p_\perp dp_\perp = E dE$, we obtain the pressure from the reflections of a incoming particle, $X$, as
\bea
\mathcal{P}_X^{r} (v)& \simeq& \frac{g_X}{4\pi^2}
\int_{-M_X}^{0} d p_z
\int_{|p_z|}^{\infty}  d E \,\,
 E (2 p_z) 
 f_{\rm incoming}.
\label{Pres:ref}
\eea 
The pressures from the transmission of incoming and outgoing $X$ is similarly given by 
\bea
\mathcal{P}_X^{t+} (v)& \simeq&\frac{g_X}{4\pi^2}
\int_{-\infty}^{-M_X} d p_z
\int_{|p_z|}^{\infty}  d E \,\,
 E \l( p_z + \sqrt{p_z^2 -M_X^2} \r) f_{\rm incoming}, 
 \\
 \mathcal{P}_X^{t-} (v)& \simeq&\frac{g_X}{4\pi^2}
\int_{0}^{\infty} d p_z
\int_{\sqrt{p_z^2+M_X^2}}^{\infty}  d E \,\,
 E \l(p_z - \sqrt{p_z^2 +M_X^2}\r) 
 f_{\rm outgoing} \simeq 0, 
\label{Pres:tr}
\eea 
where $M_X$ is the mass of the particle $X$ inside the bubble.
We numerically check that $\mathcal{P}^{t-}$ estimated with a boosted thermal distribution at $T=T_{\rm nuc}$ is indeed negligible compared to $\mathcal{P}^r$ and $\mathcal{P}^{t+}$.

\begin{figure}
      \centering
      \includegraphics[width=0.7\textwidth]{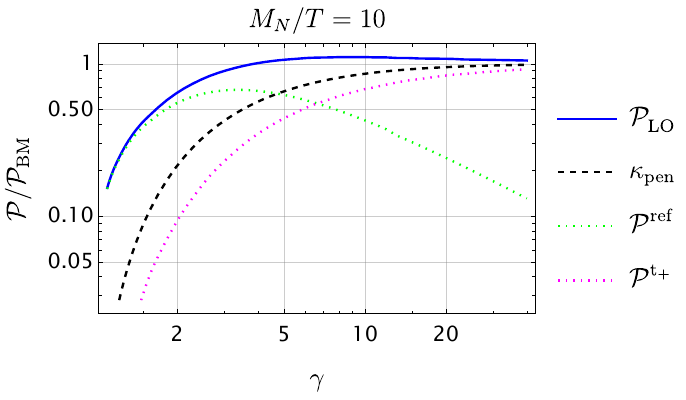}
      \caption{Different contribution to the pressure as a function of $\gamma$ normalised by the \emph{Bodeker-Moore} pressure shown in Eq.\eqref{eq:BPpres} and $M_N/T=10$ is taken. The dotted lines show the contribution from the reflected (green) and the transmitted (magenta) particles. Those two contributions add up to the full pressure in thick blue. The dashed black line on the other hand is the fraction $\kappa_{\rm pen}$ of particles entering the bubble as a function of $\gamma$.
      }
      \label{fig:int}
\end{figure}

Fig.\,\ref{fig:int} shows the pressure coming from the fermions, $N$, as a function of $\gamma$ for $M_N/T=10$.
We normalize the pressure by \cite{Bodeker:2009qy}
\bea 
\mathcal{P}_{\rm BM} \equiv \frac{1}{48} \sum_i g_i n_i M_i^2 T^2,
\label{eq:BPpres}
\eea 
where particle $i$ has $g_i$ degrees of freedom, and $n_i=1(2)$ for fermions (bosons). 
Note that, an estimation with a na{\"i}ve $\gamma^2$ approximation can lead to an incorrect conclusion.

Then, we estimate $\kappa_{\rm pen}$ as
\bea
\kappa_{\rm pen}=
\frac
{
    \int_{p_z<-M_N} \, d^3 p \, f_{\rm incoming}
}
{   
    \int_{p_z<0} d^3 p \, f_{\rm incoming}
} ,
\label{eq:kappa_pen}
\eea
where $v$ inside the expression of $f_{\rm incoming}$ is fixed by
\bea
\Delta V +\mathcal{P}_{\rm LO}(v_w)= 0 .
\eea
Here $\mathcal{P}_{\rm LO}(v_w) = \sum_X (\mathcal{P}_X^{r} (v_w)+\mathcal{P}_X^{t+} (v_w)+\mathcal{P}_X^{t-} (v_w))$ is the total leading order (LO) pressure, which includes the pressure coming from all the particles that are massive in the broken phase.
The next-to-leading order (NLO) contribution to the pressure (from $1\to 2$ processes) is not important in this context since the momentum transfer of ${\cal P}_{\rm NLO}$ is through penetrations, i.e. $\kappa_{\rm pen}\simeq 1$ when ${\cal P}_{\rm NLO}$ is important.
However, ${\cal P}_{\rm NLO}$ will be important in the study of gravitational waves since it can significantly change the energy budget of the universe during the PT.
We postpone its discussion until Section.\,\ref{sec:GW}.

\subsection{Leptogenesis inside bubbles}
\label{subsec:leptoinbubble}

Utilising $\kappa_{\rm pen}$, we solve the Boltzmann equations to evolve the initial, non-thermal number densities of $N$ which penetrate the bubble, and calculate the generated lepton asymmetry. 
In what follows, we will assume that the heavy neutrinos $N$ are in kinetic equilibrium with the SM thermal bath \emph{also inside the bubble}, thanks to the efficient rate for $\phi N \to \phi N$ via $N$ mediation. This will permit us to use integrated Boltzmann equations. 
The Boltzmann equation in this procedure can be written as
\begin{align}
\dot{n}_{N_I}+3H n_{N_I} &= 
- \sum_{A,B} \left(
        2 \langle \sigma v \rangle_{N_I N_I \to AB} \, n_{N_I}^2
        - 2 \langle \sigma v \rangle_{AB \to N_I N_I} \, n_{A} n_B
    \right)
- \Gamma_D(N_I) n_{N_I},
\label{eq:Boltzmann_nNI}
\\
\dot{n}_{B-L} +3H n_{B-L}&= 
- \sum_I \epsilon_I \Gamma_{D}(N_I)  n_{N_I} +\text{(wash-out)},
\label{eq:Boltzmann_nB-L}
\end{align}
where $\Gamma_D(N_I)$ is the total decay rate of $N_I$ and 
$\epsilon_I$ is the CP-violating parameter for $N_I$ defined by
\bea
\epsilon_{I}
\equiv
\frac
{\Gamma(N_I \to H L)
-\Gamma(N_I \to \bar H \bar L)}
{\Gamma(N_I \to H L)
+\Gamma(N_I \to \bar H \bar L)},
\eea
where $\bar L$ denotes the anti-particle of $L$.
The initial population of $N_I$ within the bubbles will be given by its massless equilibrium distribution scaled by a factor of $\kappa_{\rm pen}$:
\bea
n_{N_I}^{(0)} = 
\kappa_{\rm pen} \,
\frac{2\cdot \frac{3}{4}\cdot \zeta(3)}{\pi^2}T_{\rm nuc}^3.
\label{eq:kappa-n}
\eea

The RHNs will decay before the onset of bubble collisions, i.e. $\Gamma_D(N_I) > (\Delta t_{\rm PT})^{-1}$, where $\Delta t_{\rm PT}$ is the duration of the PT. 
To check this, it suffices to show that the lifetime of $N_I$ is shorter than $(\Delta t)_{PT} \sim (\beta_{\rm PT} H)^{-1}$,
\bea 
\frac{t_{\rm PT}}{t_{N \to HL}} = \frac{Y_D^2 M_N/8\pi }{H_{\rm reh} \beta_{\rm PT} } 
\sim
10 
\left( 
    \frac{M_N/T_{\rm reh}}{5} 
\right)^2
\left( 
    \frac{m_\nu}{0.05\,{\rm eV}} 
\right)
\left( 
    \frac{100}{\beta_{\rm PT}} 
\right),
\eea 
where $H^2_{\rm reh}\simeq 8\pi \rho_{\rm pl}(T_{\rm reh})/3M_{\rm Pl}^2$, $M_{\rm Pl}=1.2\times10^{19}\,{\rm GeV}$, and $\rho_{\rm pl}(T)=\frac{\pi^2}{30}g_* T^4$ is the radiation energy density at $T$ with effective relativistic degrees of freedom $g_* \simeq \mathcal{O}(100)$. 
Here, we take the effective neutrino mass around the atmospheric neutrino mass scale since $Y_D^2$ should be from the largest decay rate amongst the three (roughly) degenerate RHNs (see also Eq.\,\eqref{eq:YDsquaredparam0} and \eqref{eq:YDsquaredparam}); if one of RHNs decays within the duration of PT, all three RHNs effectively decay since they maintain chemical equilibrium via the efficient $N_I N_I \leftrightarrow N_J N_J$ processes, as we discuss below. Regardless, $t_{\rm PT}/t_{N\to HL}>1$ is satisfied even for the solar neutrino mass scale: $m_\nu \simeq 0.01\,\eV$.

There are various reactions relevant to Eqs.\,\eqref{eq:Boltzmann_nNI} and~\eqref{eq:Boltzmann_nB-L}.
At a minimum, the reaction rate of processes involving $\phi$ must be sizable since we need $y\sim \mathcal{O}(1)$ to ensure that $M_N/T_{\rm nuc} \sim y\,  v_\phi/\sqrt{2}T_{\rm nuc}$ is large enough to achieve washout suppression.
We highlight the two most important processes involving $\phi$ as follows:
\begin{itemize}
    \item $N_I N_I \to \phi \phi$ (model-independent)\,\cite{Shuve:2017jgj}:
The cross-section is given by
\bea 
\label{eq:sigma_dep}
\sigma_{N_I N_I \to \phi \phi}(s) \simeq\frac{3y_I^4}{128\pi} \frac{\sqrt{s-4M_I^2}}{M_I^3} +\mathcal{O}(s-4M_I^2)^{3/2},
    \eea 
where $N_I$ at this stage should be non-relativistic since $M_I/T_{\rm nuc} \sim \mathcal{O}(10)$. 
The reverse process is assumed negligible since a flat potential is required to achieve a large $M_I/T_{\rm nuc}$. This implies a smaller curvature at the minimum, i.e. light $\phi$, which, at a temperature of order $T_{\rm nuc}$, does not have enough energy to annihilate into $N_I$.
\\
With $v = \sqrt{1-4M_I^2/s}$, we estimate the thermally-averaged annihilation cross section, $\langle \sigma v \rangle$, as
\beq 
\langle \sigma_{N_I N_I \to \phi \phi} v\rangle = 
\frac{1}{16M_I^4 T K_2^2(M_I/T)}
\int_{4M_I^2}^{\infty}
ds \,
 s^{3/2} \, (1-4M_I^2/s) \, K_1(\sqrt{s}/T) \,
\sigma_{N_IN_I\to \phi\phi},
\label{eq:therm_av}
\eeq 
where we have included a factor of $1/2$ to account for the initial state phase space.
Plugging Eq.~\eqref{eq:sigma_dep} into Eq.~\eqref{eq:therm_av}, we obtain 
\bea
\langle \sigma_{N_IN_I \to \phi \phi} v\rangle &\simeq & 
 \frac{9 y_I^4 T}{128 \pi M_I^3} 
 \simeq \frac{9 \sqrt{2} y_I T}{64 \pi v_\phi^3} .
\eea
Following Appendix~D of~\cite{Shuve:2017jgj} we also verify that, in the case of an ungauged $U(1)_{B-L}$, the additional $2\leftrightarrow 2$ scattering process of $NN$ into two majorons is subdominant compared to $NN \rightarrow \phi\phi$ due to the derivative interactions involved in the regime where $M_N > T_{\rm reh}$.
\item $N_I N_I \leftrightarrow N_J N_J$ (model-independent):
Using FeynCalc\,\cite{Mertig:1990an,Shtabovenko:2016sxi,Shtabovenko:2020gxv}, we obtain the following cross section,
\bea 
\sigma_{N_I N_I \leftrightarrow N_J N_J} (s) \simeq \frac{y_I^2y_J^2  (4M_N^4-2M_N^2s+s^2)}{48\pi s^3}
\eea 
which gives the thermally averaged cross-section of the form
\bea
\langle \sigma_{N_I N_I \leftrightarrow N_J N_J} v \rangle  
 \simeq \frac{3y_I^2y_J^2  }{768 \pi M_N^2} .
\eea
This is an s-wave channel process where the velocity dependence is canceled between the initial and final states as $M_I \simeq M_J$.
\end{itemize}

While the process $N_I N_I \to \phi \phi$ depletes the population of RHNs which can decay, $N_I N_I \leftrightarrow N_J N_J$ does not.
As long as the flavor-changing processes remain efficient, we can assume that chemical equilibrium, $n_{N_1} \simeq n_{N_2} \simeq n_{N_3}$, is maintained which allows us to simplify the Boltzmann equations for $N_I$ by ignoring the number-conserving processes and assuming the same number densities for all $N_I$ at all $T$.

Defining $n_N \equiv \sum_I n_{N_I}\simeq 3n_{N_1} \simeq 3n_{N_2}\simeq 3n_{N_3}$ and $Y_N \equiv n_N/s$, we obtain
\begin{align}
\label{eq:Boltzmann_YN}
z H s \, Y_N'(z)
&=
-\bar \gamma_D \l( \frac{Y_N}{Y_N^{\rm (eq)}}-1 \r)
-2 
 \gamma_{NN\to \phi\phi}
   \l(  Y_N^2- \l( Y_N^{\rm (eq)}\r)^2 \r)
+ \text{(model-dependent)}, 
\\
\label{eq:Boltzmann_YB-L}
z H s \, Y_{B-L}'(z)
&=
-\epsilon_{\rm CP} \bar \gamma_D \l( \frac{Y_N}{Y_N^{\rm (eq)}} -1 \r)
- \frac{1}{2}(c_L+ c_H)\, \bar \gamma_D \frac{Y_{B-L}}{Y^{\rm (eq)}} ,
\end{align}
where $z\equiv M_N/T$, $Y_N^{\rm (eq)} = n_N^{\rm (eq)}/(\frac{2\pi^2}{45}g_* T^3)$ 
with the equilibrium number density $n_N^{\rm (eq)}$, 
$Y_N^{\rm (eq)} = (\frac{2}{\pi^2}T^3)/(\frac{2\pi^2}{45}g_* T^3)$, and
\bea
\bar \gamma_D &\equiv& 
\sum_I \gamma_{D}(N_I)
= \sum_I n_{N_I}^{\rm (eq)} \frac{K_I(z)}{K_2(z)} \Gamma_D(N_I), 
\\
\epsilon_{\rm CP} \, \bar \gamma_D &\equiv& 
 \sum_I \epsilon_I \gamma_D(N_I),
\\
 \gamma_{NN\to \phi\phi}
&\equiv& \frac{1}{9} s^2 
\sum \langle \sigma v \rangle_{N_I N_I \to \phi \phi}.
\eea
Note that the definition of $\epsilon_{\rm CP}$ allows it be factored out of Eq.\,\eqref{eq:Boltzmann_YB-L} by solving for $Y_{B-L}/\epsilon_{\rm CP}$ and reintroduced after solving the BEs.

Other possible $2\to 2$ processes that can affect the final asymmetry are ignored as justified in Appendix~\ref{App:scatterings}, where we show that these processes are subdominant.
In addition, following Refs.\,\cite{Nardi:2005hs, Abada:2006fw,Nardi:2006fx}, we neglect off-diagonal entries in the lepton-flavor structure of the BEs, and simply define the effective wash-out coefficient $c_L+c_H$ to account for flavour effects, which can be obtained by tracking which flavour-dependent reactions remain in chemical equilibrium at a given temperature. The relevant numerical values that we consider can be found in Table.\,\ref{tab:c_H}.

\begin{table}[t]
    \centering
    \begin{tabular}{|c|c|c|c|}
         \hline
      Temperature (GeV) & $c_L$   & $c_H$ & $c_H+c_L$ \\
      \hline
      $10^{11-12}$ & $\frac{6}{35}$   & $\frac{95}{460}$ & $\sim 0.38$ \\
      $10^{8-11}$ & $\frac{5}{53}$   & $\frac{47}{358}$ & $\sim 0.22$ \\
      $ \ll 10^{8}$ & $\frac{7}{79}$   & $\frac{8}{79}$ & $\sim 0.19$ \\
        \hline
    \end{tabular}
    \caption{$c_H+c_L$ for different relevant temperatures.}
\label{tab:c_H}
\end{table}

The decay width, $\bar \gamma_D$, is simply given by
\bea 
\bar \gamma_{D} \simeq  \frac{3}{4}\frac{g_N M_N^3}{2\pi^2 z} K_1(z) \Gamma_D \quad\mbox{with}\quad  \Gamma_D= \frac{Y_D^2M_N}{4\pi g_N},
\label{eq:rate_ID}
\eea
and we fix
% \bea 
% Y_D^2 \equiv \sum_I \left((Y_D)^\dagger Y_D\right)_{II} \gtrsim \frac{2}{v_h^2} M_N m_\nu \simeq 1.6 \times 10^{-6} \left(\frac{M_N}{10^9\text{ GeV}}\right).
% \label{eq:YDsquaredparam}
% \eea 
\begin{align}
Y_D^2 \equiv \sum_I\left( (Y_D)^\dagger Y_D\right)_{II} &= 8\pi \frac{\Gamma_{N_I}}{M_{N_I}} = \frac{2}{v_{\text{EW}}^2} \sum_I M_{N_I} \sum_i m_{\nu_i} \left| R_{Ii} \right|^2
\label{eq:YDsquaredparam0}
\end{align}
To obtain the right-hand side, we have employed the Casas-Ibarra paramaterisation~\cite{Casas:2001sr} of $Y_D$. 
In Section~\ref{sec:results} we will specialize to a classically scale-invariant potential which will require all three RHNs masses to be of the same order to prevent thermalization of the lightest RHNs within the bubbles. In such a case the above equation further simplifies:
\begin{align}
Y_D^2 &\overset{M_I \simeq M_J}{\simeq} \frac{2M_N m_{\nu_3}}{v_{\text{EW}}^2}\left( \frac{m_{\nu_1}}{m_{\nu_3}}\sum_i \left|R_{i1}\right|^2 + \frac{m_{\nu_2}}{m_{\nu_3}}\sum_i \left|R_{i2}\right|^2  + \sum_i \left|R_{i3}\right|^2 \right)\nonumber\\
&\quad\,\geq \quad\frac{2M_N}{v_{\text{EW}}^2} m_{\nu_3}
\label{eq:YDsquaredparam}
\end{align}
where, for concreteness, we have assumed a normal ordering with the hierarchy: $m_{\nu_3} \simeq 0.05\text{ eV} \gg m_{\nu_{1,2}}$. 
We stress that moving from the first line to the inequality in the final line above requires no approximation and is a rigorous lower bound in the case of exactly degenerate RHNs due to the positivity of each individual term. The above equation therefore implies that the atmospheric neutrino mass scale ($\sim 0.05\text{ eV}$) will always enter into the expression due to orthogonality of $R$ in the case of three similar mass scale RHNs. However if some different scalar potential would allow for a hierarchical spectrum of RHNs, $Y_D^2$ will no longer be the relevant quantity within the BEs but $(Y_D)^2_{11}$. In what follows we take the lower-bound for $Y_D^2$ implied by Eq.~\eqref{eq:YDsquaredparam} in our numerical estimates. We typically find that for a given choice of parameters, assuming $M_I \simeq M_J$, that $Y_D^2$ is a factor of $\mathcal{O}(5-10)$ times larger than this lower-bound. Larger values of $\bar \gamma_D$ can only increase the final net asymmetry generated, therefore our numerical results serve as a conservative lower-bound on the enhancement that can be obtained through bubble dynamics.

% \begin{sloppypar}
From Eqs.~\eqref{eq:rate_ID} and~\eqref{eq:YDsquaredparam}, $\Gamma_D \propto M_N^2$ whereas 
\bea
\Gamma_{NN\to \phi\phi} = \langle \sigma_{NN\to \phi\phi} v\rangle n_N\sim y^4 \left( \frac{T}{M_N} \right)^4 M_N.
\eea
The ratio $M_N/T$ will be fixed by phase transition properties at $T_{\rm nuc}$, therefore the relative size of the annihilation rate compared to $\Gamma_D$ will grow as the value of $M_N$ decreases. For fixed choices of $M_N/T$, there will be a lower-bound on the size of $M_N$ where bubble-assisted leptogenesis will provide an enhancement in the asymmetry. For smaller values of $M_N$, the depletion from annihilations will dominate, suppressing the asymmetry but for larger values of $M_N$, the desired enhancement will occur.
% \end{sloppypar}

\hspace{0.5cm}

The model-dependent processes in Eq.\,\eqref{eq:Boltzmann_YN} which we account for are as follows:
\begin{itemize}
\item $N_I N_I \to ff$

(model-dependent):
When gauging $U(1)_{B-L}$, the biggest obstacle comes from $N_I N_I \to ff$ annihilations where $f$ corresponds to the SM fermions. This process is important since it is not Boltzmann suppressed.
Using Ref.\,\cite{FileviezPerez:2021hbc}, we obtain
\bea
\sigma_{N_I N_I \to f_i f_i} &\simeq &
\frac{(Q^{B-L}_i)^2 g_{B-L}^4}{12\pi}
\frac{\sqrt{s-4M_I^2}}{ M_I(4M_I^2-M_A^2)^2}
+O\l((s-4M_I^2)^{3/2} \r),
\\
\label{eq:annih_NNff}
\langle \sigma v \rangle_{N_I N_I \to ff}
&\simeq&
\sum_{i=SM}
\frac{(Q^{B-L}_i)^2 g_{B-L}^4 }{4\pi}
\frac{M_I T}{(M_A^2-4M_I^2)^2},
\eea
where 
$\sum_{i=SM} 3(Q^{B-L}_i)^2 
= \frac{1}{9}(2\times 3\times 3+3\times 3+3\times 3)
+1^2 (2\times 3+3)=13$. Note that the cross section must be properly regulated as $M_A \rightarrow 2 M_N$ as in Ref.\,\cite{FileviezPerez:2021hbc}.

\item $ss \leftrightarrow N_I N_I$, $A_\mu A_\mu \to N_I N_I$
(model-dependent):
Generally, there will be additional bosonic fields, such as the scalar $s$ or the $B-L$ gauge boson $A_\mu$, required to make the PT first-order.
In order to provide a sizable contribution to the thermal potential, such a field requires a large mixed quartic coupling or gauge coupling with $\phi$.
Therefore, it is natural to assume that these bosonic degrees of freedom are heavier than $N_I$, which is the case in our numerical examples in Section~\ref{sec:results}, which implies the processes $NN\to ss$ or $NN\to AA$ are Boltzmann suppressed. Of course, the inverse of these processes will therefore not be suppressed. Accounting for these inverse processes in Eq.\,\eqref{eq:Boltzmann_YN} can only further enhance the final asymmetry as the population of $N$ which can decay will increase. However, we do not expect this to be a sizeable effect, due to the other possible annihilation channels of $s$ or $A_\mu$, so we do not include them for numerical convenience.
\end{itemize}

With all relevant model-dependent depletion processes included, we solve the Boltzmann equations of Eqs.~\eqref{eq:Boltzmann_YN} and \eqref{eq:Boltzmann_YB-L} with initial boundary conditions
\bea
Y_N(z_{\rm nuc}) = \frac{3n_{N_I}^{(0)}}{s(T_{\rm nuc})}, 
\quad 
Y_{B-L}(z_{\rm nuc})=0,
\quad
z_{\rm nuc} = \frac{M_N}{T_{\rm nuc}}.
\label{eq:initial_condition1}
\eea
Here, $z \in [z_{\rm nuc}, \, z_{\rm col}]$ and $z_{\rm col} \sim e^{H \Delta t_{\rm PT}} z_{\rm nuc} \sim 1.1 z_{\rm nuc}$. This corresponds to evaluating the total asymmetry generated at the onset of bubble nucleation, $z_{\rm nuc}$, and evaluated up to (slightly before) the temperature at which the bubbles collide, $z_{\rm col}$. Note that the definition of $n_{N_I}^{(0)}$ in Eq.~\eqref{eq:kappa-n} includes the factor of $\kappa_{\rm pen}$ which accounts for the number of RHNs which penetrate the expanding bubbles. 

The final abundances of $Y_N$ and $Y_{B-L}$ just before reheating from bubble collisions occurs is therefore given by
\bea
\tilde Y_N \equiv Y_N(z_{\rm col}) ,
\quad
\tilde Y_{B-L} \equiv Y_{B-L}(z_{\rm col}) .
\eea
These values are then used as initial conditions in the evaluation of $Y_N$ and $Y_{B-L}$ after the FOPT ends, as explained below.

\vspace{0.5cm}

\subsection{Wash-out process after bubble collisions}
The end of the phase transition occurs once the bubbles have collided, which results in a temperature increase from $T_{\rm nuc} \to T_{\rm reh}$.
A detailed modelling of reheating may affect the previous estimation, but we do not expect sizeable changes compared to our na{\"i}ve assumptions.
If the bubble wall runs away, the energy budget during the bubble expansion is mostly dominated by the kinetic energy of the bubble wall, i.e. the scalar configuration.
A collision of the scalar configuration produces, as a first step, a dominant population of $\phi$, which are much lighter than the RHNs. The annihilation and decay of the $\phi$ population reheats the universe.
The reheating process induced by the decay of $\phi$ into SM particles will not affect the number density of $N$ beyond the overall dilution factor (see Appendix~\ref{App:phi_decay} for the case where $\phi$ couples to the SM Higgs).
If the bubble wall does not run away, the shockwave formed around the bubble wall already has an increased temperature $\sim T_{\rm reh}$ since the boost factor of the bubble wall velocity is still order one. 
Therefore, there should not be a procedure that drastically changes the number density of $N_I$ beyond the aforementioned dilution factor.

We quantify the effect of bubble collisions by solving the Boltzmann equations \eqref{eq:Boltzmann_YN} and \eqref{eq:Boltzmann_YB-L}
a second time, with new initial conditions for $Y_{N}$ and $Y_{B-L}$:
\bea
Y_{N}\left( z_{\rm reh} \right)  
= \tilde Y_{N} \left( \frac{T_{\rm nuc}}{T_{\rm reh}} \right)^3,
\quad
Y_{B-L}\left( z_{\rm reh} \right)
= \tilde Y_{B-L} \left( \frac{T_{\rm nuc}}{T_{\rm reh}} \right)^3,
\quad
z_{\rm reh}=\frac{M_N}{T_{\rm reh} },
\label{eq:initial_condition2}
\eea
for $z\in [z_{\rm reh}, \, z_f]$. We choose $z_f \gg 1$ such that the final asymmetry no longer changes for larger $z$, where $\tilde Y_{N}$ and $\tilde Y_{B-L}$ were obtained from the previous step.
Then, the final baryon abundance is taken as
\bea
Y_{B} = \kappa_{\rm sph} \,Y_{B-L}(z_f),
\eea
where $\kappa_{\rm sph} = 28/79$ is the usual weak sphaleron conversion factor.

\section{Numerical results}
\label{sec:results}

\subsection{
 Classically scale-invariant models for a FOPT
}
\label{sec:model}
As the simplest example of strong FOPT, let us consider a classically scale-invariant potential of $\Phi$, which is equivalent to the conformal symmetry at the classical level\cite{PhysRevD.2.753, COLEMAN1971552, Agashe:2019lhy},
\bea
V_0(\phi) = \lambda(\mu) |\Phi|^4 = \frac{\lambda(\mu)}{4} \phi^4,
\eea
where $\phi$ is the real part of $\Phi = \frac{1}{\sqrt{2}}(\phi +i a)$ that acquires vev.

To develop a nonzero vev of $\phi$ and make the potential bounded from below, it is necessary for the $\phi^4 \log \phi$ term of the potential generated from the quantum corrections of~Eq.~\eqref{eq:CW} to be positive.
Since $N_I$ produces a negative contribution, and is necessary for leptogenesis, new bosonic degrees of freedom that couple to $\Phi$ in a sizable way are required.
We consider two possible options:
\begin{itemize}
    \item{Scalar catalyzed (SC): Spectator scalar field $s$
    \\
    The gauge-singlet real scalar field, $s$, contributes to the effective potential via $m_s^2(\phi)=\lambda_{s\phi } \phi^2$, i.e. $-\Delta {\cal L} = \frac{\lambda_{s\phi}}{2} s^2 |\Phi|^2$.
    }
    \item{Gauge boson catalyzed (GBC): Gauged $U(1)_{B-L}$
    \\
    The $B-L$ gauge boson, $A_\mu$, contributes to the effective potential via $m_A(\phi)=2g_{B-L} \phi$, where $g_{B-L}$ is the gauge coupling and we assume $\Phi$ has a charge of $2$ to allow for Eq.\,\eqref{Eq:L_RHN}.
    }
\end{itemize}
One may possibly consider combining both models simultaneously as in Ref.\,\cite{Huang:2022vkf}. In this work, we consider them separately for simplicity\,\footnote{
In Ref.\,\cite{Huang:2022vkf}, it was argued that $M_N/T_{\rm reh}$ cannot be large enough for bubble-assisted leptogenesis in the gauge boson catalyzed case without introducing additional scalar fields. 
We find that this is not the case as: i) $M_N/T_{\rm reh}$ required to avoid a strong wash-out of $Y_{B-L}$ is only $\gsim 7$, and ii) the effective potential difference is loop suppressed, so the argument made in Ref.\,\cite{Huang:2022vkf} should be modified to $M_N/T_{\rm reh} < M_A/T_{\rm reh} \sim \text{(a few)}\times10$, where $M_A$ is the $B-L$ gauge boson mass, which cannot rule out this model.
We similarly find that depletion effects are sizeable, but by numerically solving the relevant BEs we still find a possible enhancement in the final asymmetry compared to the conventional scenario.} but for notational convenience, we write both contributions from $g_{B-L}$ and $\lambda_{s\phi}$ in some expressions.

At zero temperature, the effective potential of $\phi$ is given by
\bea
V_{\rm eff}(\phi)=V_0(\phi) + \sum_i V_{\rm CW}(m_i^2(\phi)),
\eea
where the field-dependent $\overline{\text{MS}}$ masses can be written as
\bea
  m_\phi^2 = 3\lambda \phi^2, 
  \qquad 
  m_a^2 = \lambda \phi^2,
  \qquad
  M_{N_I}^2 =  y_I^2\phi^2/2, 
  \qquad 
  m_A^2 = 4 g_{B-L}^2 \phi^2, 
  \qquad 
  m_s^2 = \lambda_{s\phi} \phi^2.
\eea
This leads to an expression of
\bea
V_{\rm eff}(\phi) = 
\frac{1}{4}
\left[\,
\lambda(\mu) + \beta_\lambda \log \phi/\mu + \delta \lambda (\lambda, y_I, \cdots) 
\, \right]
\phi^4,
\label{eq:effective_potential_in_beta}
\eea
where $\beta_\lambda\equiv d\lambda/d\ln\mu$ is the one-loop beta function\,\footnote{
We should have also included the running of $\phi$ through $\gamma$, the anomalous dimension of $\phi$; $\phi \to \phi_0 e^{\Gamma(\mu)}$ where $\Gamma(\mu)=\int_{\mu_0}^\mu \gamma(\mu') d\ln \mu'$.
However, its contribution in Eq.\,\eqref{eq:effective_potential_in_beta} is always multiplied by $\lambda(\mu)$ while  $\lambda(\mu)$ itself is \emph{numerically} one-loop order unless we take a pathological RG scale. 
Therefore, we simply ignore its contribution in this paper.
}
for $\lambda$, and $\delta \lambda(\lambda, y_I, \cdots)$ includes a function of coupling constants and does not have any explicit dependence on $\phi$.
For instance, in our model, we have
\bea
16\pi^2 \delta \lambda(\lambda, y_I, \cdots) &=& 
\lambda^2 (10 \log \lambda +9 \log 3-15)
-\sum_I \frac{1}{2}y_I^4(\log y_I^2/2 -3/2) \nn 
&+&12 g_{B-L}^4 (\log 4g_{B-L}^2 -5/6)
+ \lambda_{s\phi}^2 (\log\lambda_{s\phi}-3/2).
\eea
The $\mu$ dependence in $\delta \lambda(\lambda, y_I, \dots)$ comes only via the RG running of coupling constants, so its effect on $V_{\rm eff}$ is at two-loop order.
This would be canceled if the effective potential was calculated at two-loop order but we calculate only up to one-loop order which implies that $V_{\rm eff}(\phi)$ is $\mu$-independent only up to one-loop order.

One can use the minimization condition $V_{\rm eff}'(\phi)=0$ and express $v_\phi\equiv \langle \phi \rangle$ in terms of coupling constant defined at a given RG scale $\mu$.
For example, if we take $\mu= v_\phi$, one can find the minimization condition $\lambda(v_\phi)+\delta \lambda=-\beta_\lambda/4$.

We fix the RG scale at $\mu =\mu_*$, which is defined as $\lambda(\mu_*)\equiv 0$ for simplicity.
This allows us to ignore all the $\lambda(\mu_*)$ contributions.
For $\mu=\mu_*$, we obtain
\bea
v_\phi \equiv \langle \phi \rangle
= e^{-(\delta \lambda/\beta_\lambda +1/4)}\mu_* ,
\eea
where
\bea
\hspace{-1cm}
\beta_\lambda |_{\mu=\mu^*}&=& \frac{1}{16\pi^2}
\left(
    -\sum_I  y_I^4
    +96 g_{B-L}^4
    +2 \lambda_{s\phi}^2
\right),
\\
\delta \lambda |_{\mu=\mu_*}&=& \frac{1}{16\pi^2}
\left(
    -\sum_I \frac{1}{2}y_I^4(\log y_I^2/2 -3/2)  
    +12 g_{B-L}^4 (\log 4g_{B-L}^2 -5/6)
    + \lambda_{s\phi}^2 (\log\lambda_{s\phi}-3/2).
\right).
\nn
\eea
The exponent ${(\delta \lambda/\beta_\lambda +1/4)}$ is not large since both $\delta \lambda$ and $\beta_\lambda$ are at one-loop order.

The temperature dependence of the potential can be obtained using Eq.\,\eqref{eq:VT} with the thermal mass corrections to the bosonic states given by 
  \bea
  \Pi_{A_T} = 0, 
  \qquad \Pi_{A_L} = g_{B-L}^2\frac{T^2}{3}, \qquad \Pi_s = \frac{\lambda_{s\phi} T^2}{3},
  \eea
where $A_{T (L)}$ corresponds to the transverse (longitudinal) components of $A_\mu$.
Here, $\Pi_\phi$ and $\Pi_a$ are omitted since their contributions are suppressed: $\lambda(\mu_*)=0$.

\begin{figure}[t]
    \centering
    \includegraphics[width = 0.49\textwidth]{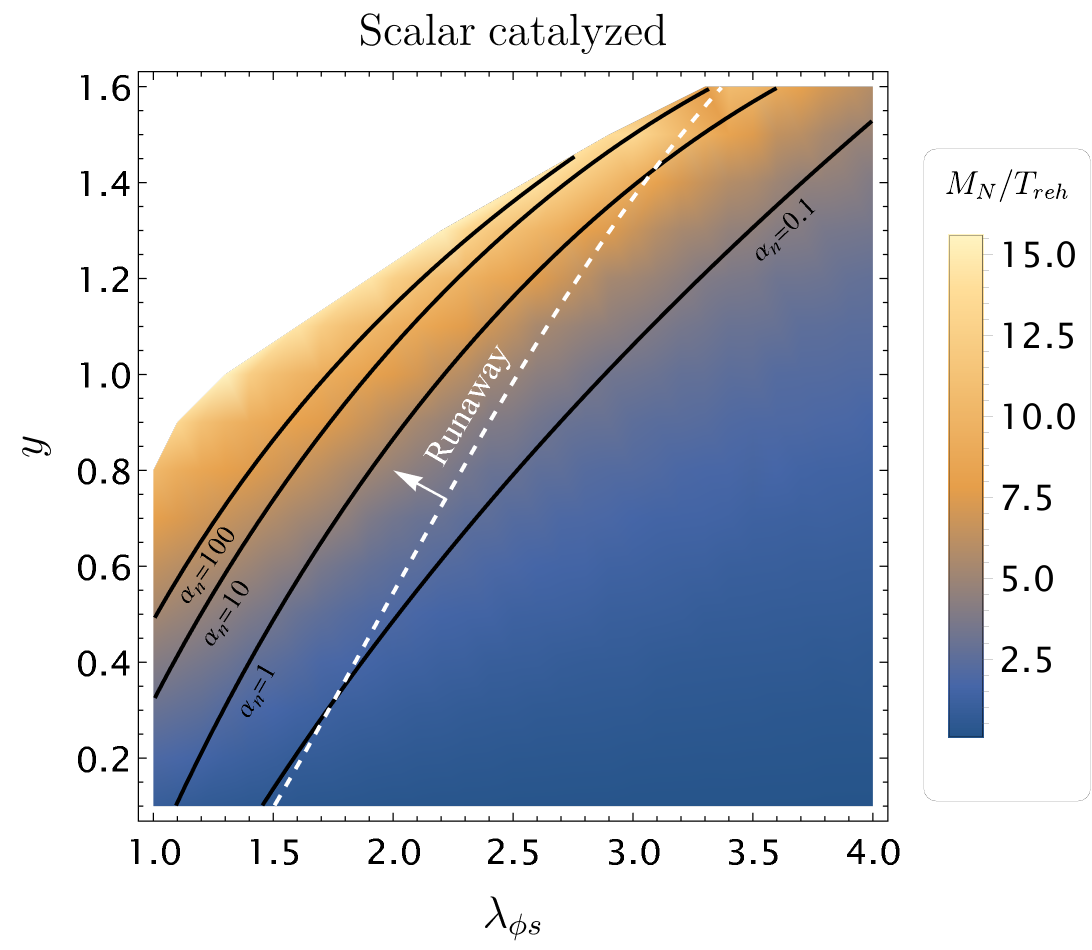}
    \includegraphics[width = 0.48\textwidth]{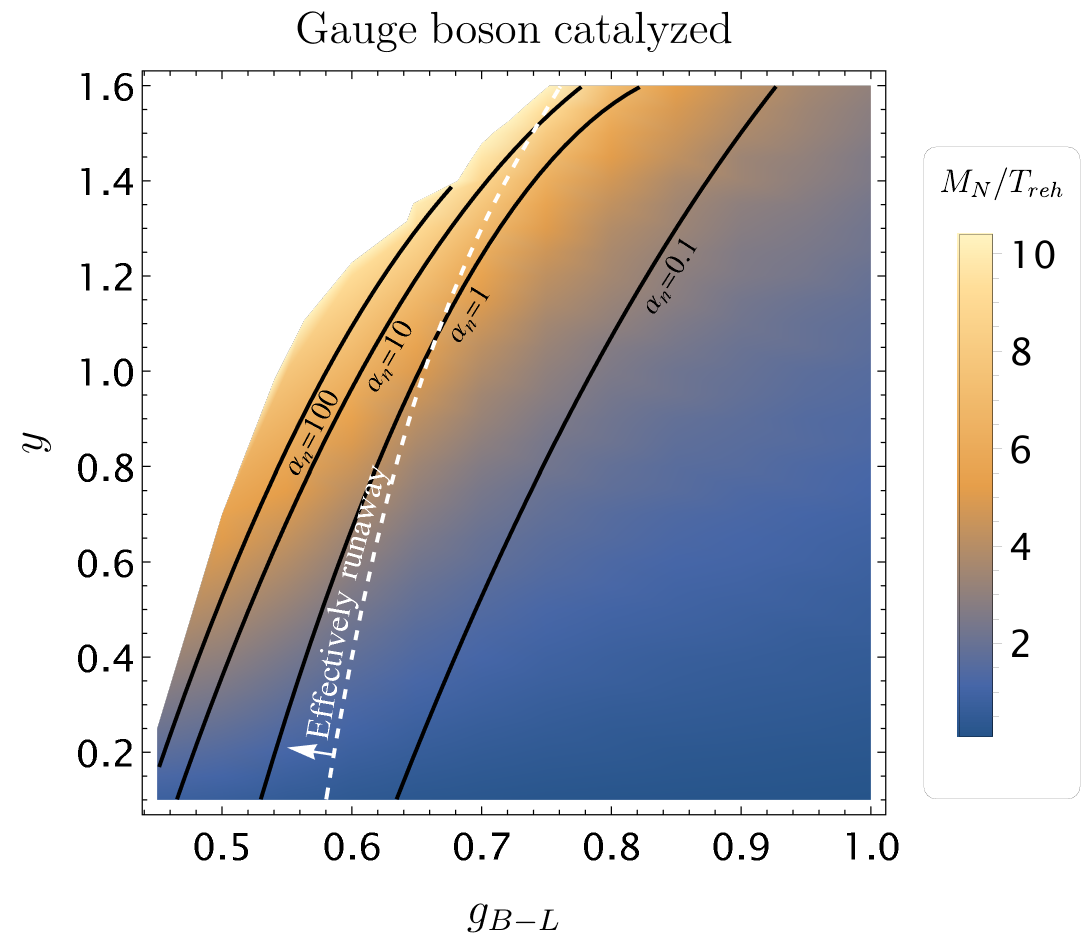}
    \caption{Parameter scan for the scalar and gauge boson catalyzed case. For both cases is the region of $\alpha_n \sim \mathcal{O}(1)$ 
    in the (effectively) runaway regime. }
    \label{Fig:MNoverTreh}
\end{figure}

Based on the formalism presented in Section\,\ref{sec:review}, we obtain values for $M_N/T_{\rm reh}$ and $\alpha_n$ as a function of $y=y_I$ and $\lambda_{s\phi}$ ($g_{B-L}$) in the SC (GBC) case.
Fig.\,\ref{Fig:MNoverTreh} shows the variation of $M_N/T_{\rm reh}$ together with $\alpha_n$, indicated by the black solid lines for $\alpha_n=0.1,\,1,\,10$ and $100$, in the SC (left panel) and GBC (right panel) scenarios.
To minimize the suppression from $\kappa_{\rm wash}$ (as well as $\kappa_{\rm dep}$), one needs $M_N/T_{\rm reh}\gsim 7$  as can be seen in Fig.~\ref{fig:summary}. On the other hand,  $\alpha_n$ should be kept to be order one to avoid a large dilution: $(T_{\rm nuc}/T_{\rm reh})^3 =  (1+\alpha_n)^{-3/4}$.

Therefore, we find viable parameter space compromising these conflicting effects at $y\sim \mathcal{O}(1)$ and $\lambda_{s\phi}\sim \mathcal{O}(2)$ (SC) or $g_{B-L}\sim \mathcal{O}(1)$ (GBC).
Around this region, the penetration rate $\kappa_{\rm pen}$ is close to one since the bubble wall runs away as  indicated by the white dashed curve. This is the boundary of the bubble wall running-away and reaching a terminal velocity, which is determined by using Eqs.~\eqref{Pres:ref} and~\eqref{Pres:tr}.

\subsection{Comparison with conventional leptogenesis}
The values of $M_N/T_{\rm reh}$ and $\alpha_n$ obtained from the previous section determine the initial conditions when solving the BEs before and after bubble collisions, Eqs.~\eqref{eq:initial_condition1} and~\eqref{eq:initial_condition2} respectively.

\begin{figure}[t]
      \centering
      \includegraphics[width=0.32\textwidth]{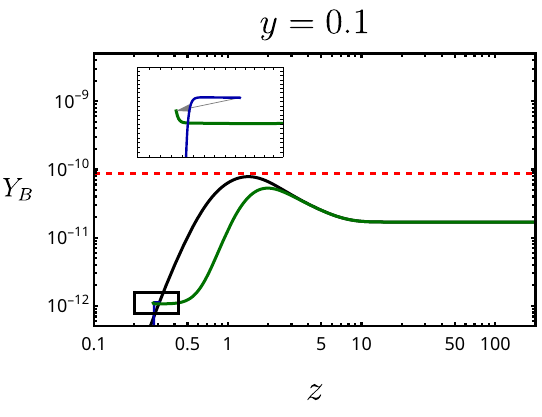}
      \includegraphics[width=0.32\textwidth]{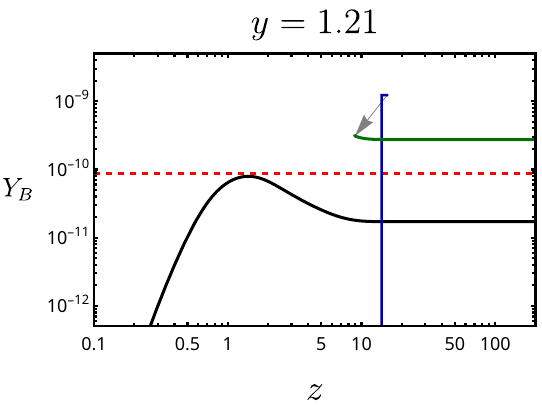}
      \includegraphics[width=0.32\textwidth]{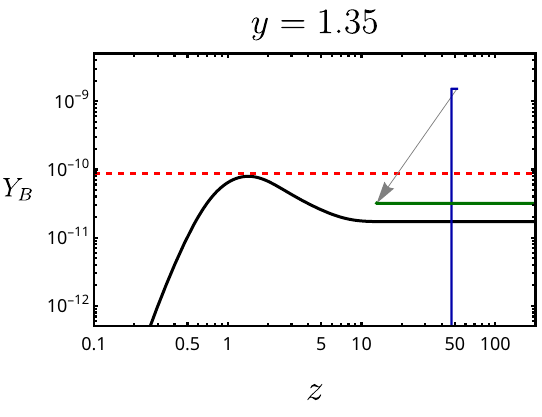}
      \caption{Numerical solution to the BEs presented in Eqs.~\eqref{eq:Boltzmann_YN} and \eqref{eq:Boltzmann_YB-L} assuming $M_N = 5\times 10^{9}$ GeV and $\lambda_{s\phi} = 2.5$ for different values of the Yukawa couplings, $y=y_I$, in the ungauged case with a real singlet scalar (SC). The black lines correspond to the usual thermal scenario (without any additional $B-L$ scalars), the blue lines to the asymmetry produced from RHN decay before bubble collision (between $z_{\rm nuc}$ and $z_{\rm col}$) and the green lines track the evolution of the asymmetry after $T_{\rm reh}$ with the required asymmetry indicated by the dashed-red line. The gray arrow indicates reheating of the bath from $z_{\rm col}$ to $z_{\rm reh}$ from bubble collisions. For small values of $y$, $M_N < T_{\rm nuc}, T_{\rm reh}$, and the RHNs thermalise within the bubbles before decaying, recovering the usual thermal scenario (plus the additional depleting interactions $N_I N_I \rightarrow \phi\phi$). For values of $y$ where $\alpha_n\sim \mathcal{O}(1)$, the condition $M_N > T_{\rm nuc}, T_{\rm reh}$ is satisfied so washout and depletion are suppressed, and an enhancement compared to the thermal scenario is observed. For larger values of $y$ however, where $\alpha_n$ grows sharply with $y$, $T_{\rm nuc}/T_{\rm reh} \ll 1$ and reheating drastically suppresses the final asymmetry generated. For concreteness, $\epsilon_{\rm CP} \simeq Y_D^2/8\pi$ was assumed in both cases.} 
     \label{fig:recreate-benchmark_YBL}
\end{figure}

As an illustrative example, Fig.~\ref{fig:recreate-benchmark_YBL} depicts the evolution of $Y_{B-L}$ in the scalar catalyzed scenario. Here, $\lambda_{s\phi} = 2.5$ and we vary the Majorana Yukawa coupling $y= y_I$. We have intentionally fixed $M_N = 5\times 10^9 \text{ GeV}$ for all choices of $y$, such that conventional strong-washout thermal leptogenesis does not generate sufficient asymmetry. The solid-blue lines denotes the evolution of $Y_{B-L}$ for bubble-assisted leptogenesis during the time scales between $z_{\rm nuc}$ and $z_{\rm col}$. After $z_{\rm col}$, the universe is reheated up to $z_{\rm reh}$ and $Y_{B-L}$ is diluted by a factor $(T_{\rm nuc}/T_{\rm reh})^3$, which we indicate by the grey arrow. The solid-green line tracks the evolution of $Y_{B-L}$ from $z_{\rm reh}$ onwards. 

We contrast these results with the conventional thermal leptogenesis scenario, denoted by the solid-black line, where: we have turned off all $\phi$-related annihilation processes (which could only reduce the asymmetry further), we have assumed a hierarchical spectrum of RHNs but assumed that $\gamma_{N_1} = \bar \gamma_D $ from Eq.~\ref{eq:rate_ID}, and used the initial conditions
\bea
Y_{N_1}(z_i) = Y_{N_1}^{\rm eq}(z_i),\qquad Y_{B-L}(z_i) =0
\eea
for $z \in [z_i,\,z_f] = [0.1,\,200]$. For both the bubble-assisted and conventional leptogenesis evolutions, we have assumed $\epsilon_{\rm CP} \simeq Y_D^2/8\pi$ for concreteness. Smaller (or larger) values may change the overall asymmetry obtained in both cases, however the relative enhancement provided from the bubble-assisted scenario should not vary significantly. Similar behaviour will be obtained in the gauged scenario where the only significant difference is the additional presence of the $N_I N_I \rightarrow f f$ depleting processes which can slightly decrease the final asymmetry within the bubble-assisted scenario.

We repeat this procedure by scanning the FOPT parameter space, and summarize our results in Fig.~\ref{fig:summary} as a function of $\alpha_n$, again assuming $M_N = 5\times 10^9\text{ GeV}$. The left and right panels correspond to the scalar catalyzed (SC) and gauge boson catalyzed (GC) scenarios respectively, and the grey bands denote the size of the enhancement compared to conventional leptogenesis. The grey band boundaries are obtained by choosing $y \in [0.85,\,1.4]$ in the scalar catalyzed case and $y \in [1.0,\,1.13]$ in the gauged case. Larger couplings lead to a larger enhancement, for a fixed $\alpha_n$, as indicated by Fig.~\ref{Fig:MNoverTreh} where the ratio $M_N/T_{\rm reh}$ grows with $y$, $\lambda_{s\phi}$ and $g_{B-L}$. The horizontal axis of Fig.~\ref{fig:summary} shows the strength of the supercooling, $\alpha_n$, which corresponds to scanning $\lambda_{s\phi}$ for the SC and $g_{B-L}$ for the GBC respectively for the given value of $y$ and $M_N$.

In Fig.~\ref{fig:summary}, we also depict the various suppression factors we have defined, $\kappa_{\rm pen}$, $\kappa_{\rm wash}$, $\kappa_{\rm dep}$ and $(T_{\rm nuc}/T_{\rm reh})^3$, by the dark red, bright red, yellow and blue dashed curves respectively. For concreteness, we fixed $y$ to the largest value displayed on each figure when estimating these suppression factors, which corresponds to the largest possible enhancement compared to the conventional scenario. The penetration coefficient, $\kappa_{\rm pen}$, remains close to unity in our chosen parameter space but begins to decrease for $\alpha_n \lsim 1$. Smaller values of $\alpha_n$ imply a large washout, $\kappa_{\rm wash}$ and $\kappa_{\rm dep}$, as the ratio $M_N/T_{\rm reh}$ decreases. Much larger values of $\alpha_n$ however, suffer from a large diluting effect: $(T_{\rm nuc}/T_{\rm reh})^3 = (1+\alpha_n)^{-3/4}$.
The enhancement is maximized around $\alpha_n \sim 5$, since $Y_B$ is proportional to the combination of all these factors: Eq.~\eqref{eq:finalasymparam}.

\begin{figure}
      \centering
      \includegraphics[width=0.45\textwidth]{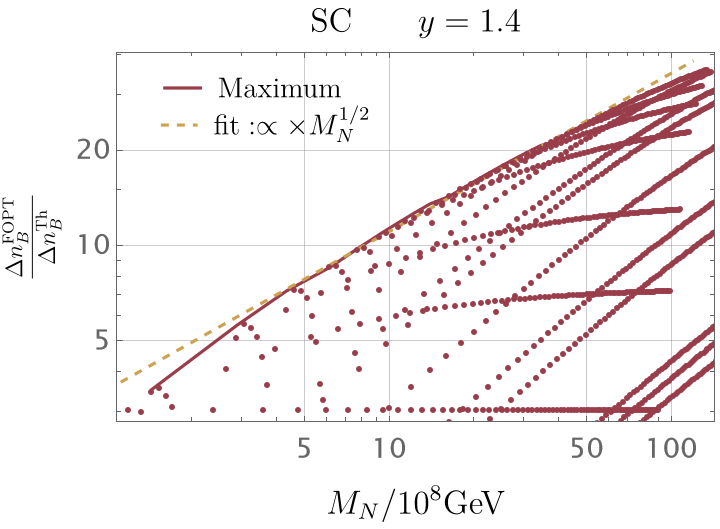}
      \includegraphics[width=0.45\textwidth]{figs/ScatPlotGBC.png}
      \caption{Scatter plot of $M_N$ versus the enhancement for $y = 1.4$ for SC (Left) and GBC (Right) cases. This scatter plot suggests that, due to stronger depletion at low $T$, the enhancement disappears around $M_N \sim 10^7$ GeV. The upper thick lines represent the maximal enhancement (at the peak) and is thus an upper bound. This upper bound for the enhancement can be roughly analytically fitted by the dashed yellow line with expression in Eq.~\eqref{eq:fit}. For fixed mass $M_N$, we scan over different values of $\alpha_n$. 
      The patterns appearing in the above figure are spurious artifacts due to the method of scanning.
      }
      \label{fig:scatter}
\end{figure}

The maximal enhancement factor can be as large as $\sim 20$ for $M_N = 5 \times 10^9\,\GeV$, however the enhancement decreases as we decrease $M_N$. As shown in Fig.~\ref{fig:scatter}, we roughly find that below $M_N \simeq 10^7 \text{ GeV}$ bubble-assisted leptogenesis cannot provide an enhancement compared to the conventional scenario. This conclusion should be insensitive to the choice of $\epsilon_{\rm CP}$.
This lower-bound arises as the $\phi$- and $A_\mu$-induced depletion processes, which we quantify through $\kappa_{\rm dep}$, grow as $M_N$ decreases, as discussed in Section~\ref{subsec:leptoinbubble}; $\Gamma_{\rm ann} \propto z_{\rm nuc}^{-4} M_N$ while $\Gamma_D \propto M_N^2$. Due to the additional depletion channels which occur for the gauged case, we observe a slightly larger depletion effect compared to the scalar catalyzed case in Figs.~\ref{fig:scatter} and~\ref{fig:summary}, as expected.

Interestingly, from Fig.~\ref{fig:scatter}, we find that the \emph{maximal enhancement} is well fitted by 
\bea 
\frac{n_B^{\rm FOPT, max}}{n_B^{\rm thermal}} \sim \bigg(\frac{M_N}{10^7 \text{GeV}}\bigg)^{1/2}.
\label{eq:fit}
\eea

\section{Gravitational waves}
\label{sec:GW}
As bubble-assisted leptogenesis relies on bubble dynamics, 
it is an important question whether gravitational waves produced during the FOPT will be detectable in future gravitational wave detectors such as LIGO\cite{LIGOScientific:2007fwp}, ET\cite{Maggiore:2019uih} and CE\cite{Evans:2021gyd}.

In general, there are three sources of gravitational waves during a FOPT; collisions of bubble walls (scalar), sound waves, and turbulence effects in the fluid 
(see Refs.\,\cite{Alanne:2019bsm, Caprini:2019egz, Caprini:2015zlo} for nice reviews and also Appendix\,\ref{App_GW} for the semi-analytic expressions we use in our estimations). 
The scalar contribution is the dominant source when the bubble walls run away while the sound wave contribution becomes important when the bubble wall has a constant velocity.
This results in qualitatively different regimes in the gravitational wave spectra between the SC case and the GBC case.
As we discussed, at the position of the enhancement peak, the plasma pressure on the wall via the $1\to1$ process cannot reach a balance with the potential difference $\Delta V$.
Meanwhile, in the GBC case, $1\to2$ processes, e.g. massless $N$ outside bubbles to massive $N$ and $A_\mu$ inside bubbles, can provide an additional source of pressure as\,\cite{Bodeker:2017cim, Gouttenoire:2021kjv}
\bea
\mathcal{P}_{\text{NLO}} \sim \frac{1}{16\pi^2} T^3\gamma_w g^3\Delta m_{\rm GB}.
\label{NLO}
\eea
Since it increases as $\gamma_w$, the bubble wall eventually reaches terminal velocity, and shock waves can be formed around the bubble wall during the expansion.
Accounting for these effects in Section~\ref{subsec:penrate} and Fig.~\ref{Fig:MNoverTreh} will not induce any appreciable change in the estimate for $\kappa_{\rm pen}$ since, although $\gamma_w$ is now finite, the terminal velocity the bubble wall reaches is so large that $\kappa_{\rm pen} \simeq 1$ as justified by Fig.~\ref{fig:int}. However, this implies that the scalar contribution to the gravitational wave spectrum is less important compared to the sound wave contribution.
The difference between these two contributions appears in the high frequency tail where the spectrum falls as $f^{-1.5}$ for the scalar contribution and $f^{-4}$ for the sound wave contribution.
On the other hand, both contributions have a peak frequency proportional to the inverse of the average value of a bubble radius at the time of percolation multiplied by the redshift factor due to the Hubble expansion until today. 
Therefore, the peak frequency is proportional to the reheating temperature $T_{\rm reh}$ where $T_{\rm reh} \sim (\Delta V)^{1/4} \propto M_N$.

\begin{figure}[t]
      \centering
      \includegraphics[width=0.45\textwidth]{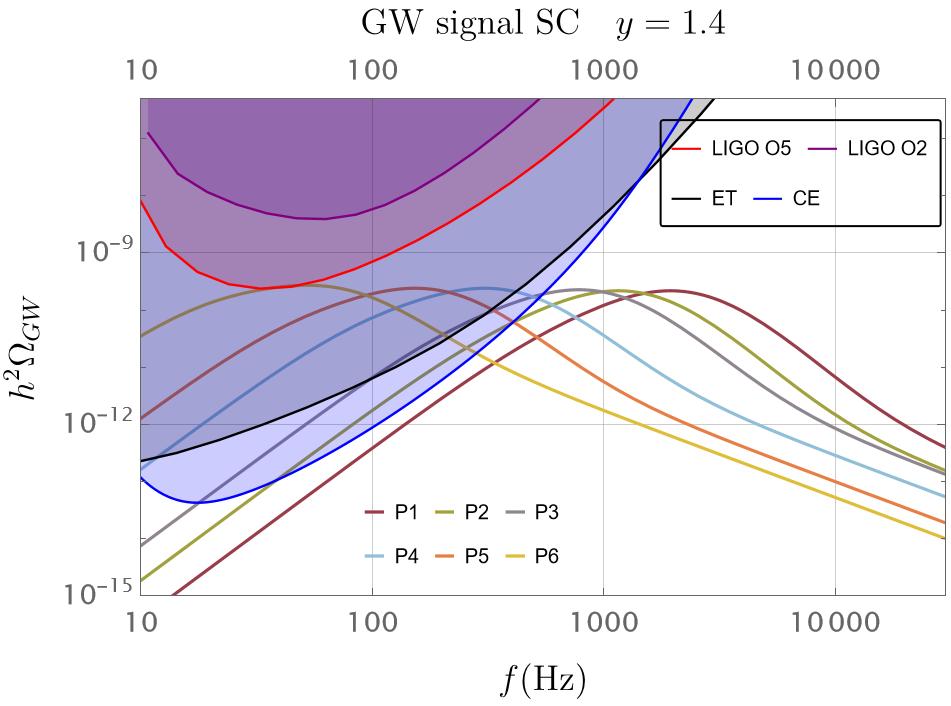}
      \includegraphics[width=0.45\textwidth]{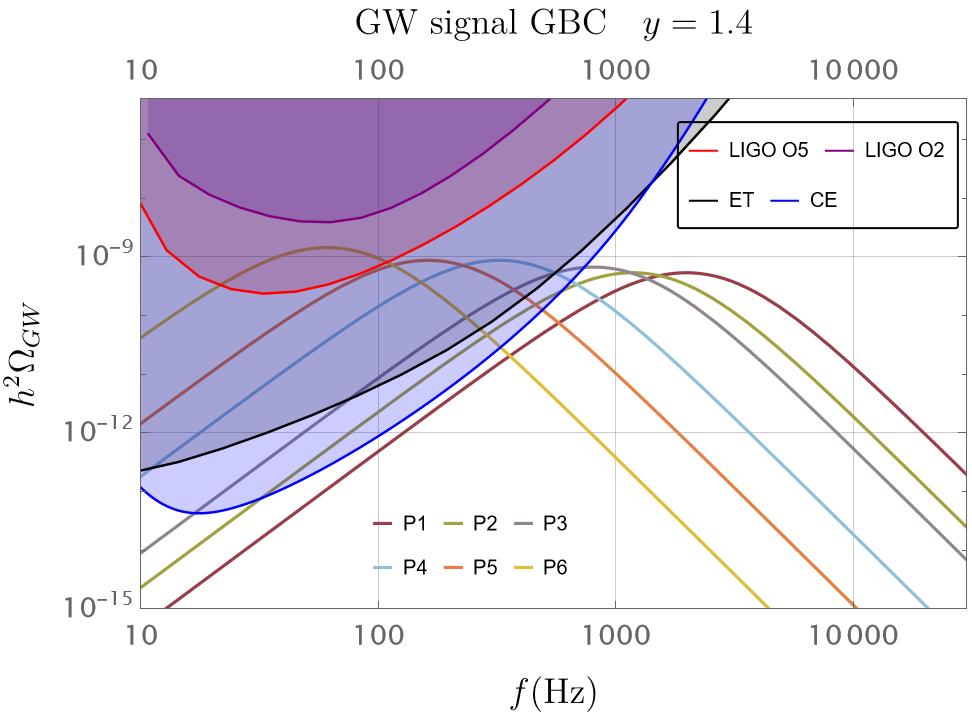}
     \caption{GW signal from leptogenesis for various benchmark points summarized in Table\,\ref{tab:BMp}.
     Future sensitivity of LIGO, ET and CE are obtained by following Refs.\,\cite{ Moore:2014lga,Aasi:2013wya,TheLIGOScientific:2014jea,Cornish:2018dyw,Graham:2017pmn,Yagi:2011yu,Yagi:2013du,Sathyaprakash:2012jk}. For the sound waves we took $\tilde \Omega_{\rm GW} \sim 0.01$\cite{Hindmarsh:2017gnf, Cutting:2019zws} for detonations.
}
     \label{fig:GW}
\end{figure}

\begin{table}[t]
    \centering
    \begin{tabular}{|c|c|c|c|c|c|c|}
    \hline
  & \multicolumn{3}{|c|}{SC} & \multicolumn{3}{|c|}{GBC} \\
 \hline
       & $ M_N/10^8 \text{GeV}  $  & $\frac{n_B^{FOPT}}{n_B^{\rm thermal}}$ & $\alpha_n$ & $ M_N/10^8 \text{GeV}  $  & $\frac{n_B^{FOPT}}{n_B^{\rm thermal}}$ & $\alpha_n$ \\
      \hline
      P1 & $62$   & $26$ & $4.4$ & $60$   & $22$ & $4.8$ \\
      P2 & $34$   & $21$ & $6$ & $37$   & $17$ & $5$ \\
      P3 & $26$   & $18$ & $6.$  & $25$   & $15$ & $9$\\
      P4 & $10$   & $12$ & $10$ & $10$   & $10$ & $9.6$ \\
      P5 & $4.2$   & $8$ & $25$ & $4$   & $5.6$ & $15$ \\
      P6 & $1.4$   & $3.5$ & $25$ & $1.4$   & $2.9$ & $33$\\
        \hline
    \end{tabular}
    \caption{Benchmark points for the GW signal in Fig.\ref{fig:GW} where all $M_{N_I} \equiv M_N$. 
    }
\label{tab:BMp}
\end{table}

In Fig.\,\ref{fig:GW}, we show the gravitational wave spectra for the SC and GBC case for various benchmark parameters of $M_N$, the final asymmetry enhancement factor (compared to thermal leptogenesis), and $\alpha_n$. Their values are summarized in Table\,\ref{tab:BMp}.
These benchmark points were chosen as they maximize the enhancement factor for each choice of $M_N$, i.e. $M_N/T_{\rm reh}\sim 8$ and $\beta_{\rm PT}\sim 50$.
Future sensitivity of LIGO, ET and CE are obtained by following Refs.\,\cite{ Moore:2014lga,Aasi:2013wya,TheLIGOScientific:2014jea,
Cornish:2018dyw,Graham:2017pmn,Yagi:2011yu,Yagi:2013du,Sathyaprakash:2012jk}.
We find a rather strong GW signal due to $\alpha_n \sim 1$ and not-too-large $\beta_{\rm PT}$, as expected from a pure CW potential \cite{DelleRose:2019pgi, Levi:2022bzt}.

As can be seen in Fig.\,\ref{fig:GW}, the gravitational waves are detectable for $M_N \lsim 10^{9}\,\GeV$. This low-mass region of $M_N$ is perhaps the most motivated region for bubble-assisted leptogenesis, as in this region thermal leptogenesis cannot generate sufficient lepton asymmetry without a tuning of parameters. Larger values of $M_N$ will move the peak frequency of GWs away from upcoming experiments, preventing detectability, however in this region thermal leptogenesis can occur without issue and bubble-assisted leptogenesis might be considered a less interesting mechanism.

As shown in the Fig.\,\ref{fig:GW}, although the bubble wall runs away in the SC case, the sound wave contribution dominates as ${\cal P}_{\rm LO}/\Delta V$ is not too suppressed. The difference between these two cases will appear in the UV tail as we discussed above, but as this is outside the sensitivity of future GW detectors\footnote{{It has been however recently claimed~\cite{Lewicki:2022pdb} that in the case of strong transition $\alpha_n \gg 1$ and fast bubbles $\gamma_w \gg 1$, the spectrum of the sound waves induced GW might be indistinguishable from the bubble collision component. The UV tail of the GW signal might change in future studies but we consider this beyond the scope of the current work.}}, these two cases will not be distinguishable at upcoming experiments.

In the case of spontaneously broken Abelian symmetries, cosmic strings can be formed and produce an appreciable spectrum of gravitational waves\cite{Dror:2019syi, Blasi:2020wpy}. However, the amplitude is typically small if $v_\phi \lesssim 10^{11}$ GeV, which is the rough region of interest we consider, where thermal leptogenesis may be insufficient without tunings. We have therefore ignored such possible sources of GWs.

\section{Conclusions}
\label{sec:conclusion}
We have investigated the size of the enhancement possible within bubble-assisted leptogenesis compared to the conventional thermal leptogenesis scenario, assuming a classically scale-invariant scalar potential as the source of the FOPT. Additionally, we have briefly explored the GW implications in this scenario, specifically focused on those produced during the FOPT itself.

Although we have a stronger departure from the thermal equilibrium compared to the conventional scenario, such that a large wash-out suppression can be avoided, there are alternative processes necessarily predicted which serve to dilute the final asymmetry yield which are not present in the conventional thermal scenario. These include the dilution due to reheating and CP-conserving depletion processes of the RHN population from $2 \rightarrow 2$ annihilation to other $B-L$ fields.
We have established a systematic approach to include all these effects for a general FOPT potential and a step-by-step description on how to evaluate the baryon yield from bubble-assisted leptogenesis.

We numerically find that the most favorable range for the mass, $M_N$, in bubble-assisted leptogenesis is roughly $10^9-10^{10}\,\GeV$ and an $\mathcal{O}(20)$ enhancement of $Y_B$ in this region compared to conventional leptogenesis.
For larger values of $M_N$, an even larger sized enhancement is possible however, conventional thermal leptogenesis can efficiently produce the required asymmetry in this mass range so bubble-assisted leptogenesis loses some of its appeal. An enhancement remains possible in the region $M_N \in [10^7,\,10^9]\,\GeV$ compared to the usual thermal scenario but is insufficient in generating the observed asymmetry, therefore a supplementary mechanism which can provide an additional enhancement of $\epsilon_{\rm CP}$ is required.

An important characteristic of bubble-assisted leptogenesis that we highlight is that the enhancement rapidly disappears when $M_N$ falls below $10^8\,\GeV$. This is because
$\Gamma_{\rm ann} \propto  M_N$ while $\Gamma_D \propto M_N^2$.
We cannot yet conclude that this is an unavoidable consequence, as we have restricted ourselves to classically scale-invariant potential so there still remains a logical possibility that a different phase transition model can provide a larger value of $z_{\rm nuc}$ such that $\Gamma_{\rm ann}$ is suppressed: $\Gamma_{\rm ann}\propto z_{\rm nuc}^{-4}$.
However, whether there exists a model such that a larger $z_{\rm nuc}$ can occur whilst simultaneously allowing for $\alpha_n \sim \mathcal{O}(1)$ is doubtful.

Bubble-assisted leptogenesis predicts gravitational wave signals with frequencies around $\mathcal{O}(10^2)$-$\mathcal{O}(10^4)\,{\rm Hz}$.
Their spectrums are mostly given by the sound wave contribution during the FOPT, both for the SC case as well as the GBC case, in the phase transition parameter space where the enhancement of $Y_B$ compared to the conventional scenario is large. The parameter range of $M_N \lsim 5\times 10^9\,\GeV$ is within the sensitivity of future gravitational wave detectors like ET, CE or LIGO O5. 
As the favorable mass range for $M_N$ in bubble-assisted leptogenesis is around $10^9\,\GeV$--$ 10^{10}\,\GeV$, where the observed baryon asymmetry can be explained with a natural choice of $\epsilon_{\rm CP}$, these upcoming detectors can probe an important parameter regime of the scenario.
Gravitational wave detectors with the ability to probe higher frequency ranges, which have been recently discussed in Refs.~\cite{Aggarwal:2020umq,Domcke:2022rgu,Chou_2017,Goryachev:2014yra,Berlin:2021txa,Berlin:2023grv}, will complement the terrestrial-based detectors, provided that an improvement in the sensitivity is possible.

\section*{Acknowledgement}

 MV and XN are supported by the ``Excellence of Science - EOS" - be.h project n.30820817, and by the the Strategic Research Program High-Energy Physics of the Vrije Universiteit Brussel and would like to thank Giulio Barni, Aleksandr Azatov, Wen Yin, Marco Drewes, Iason Baldes and Alberto Mariotti for insightful discussions and useful comments on the draft. TPD is supported by KIAS Individual Grants under Grant No. PG084101 at the Korea Institute for Advanced Study.
 This work was supported by IBS under the project code, IBS-R018-D1. We would like to thank as well the organizers of the IPMN event ``what the heck occurs when the universe boils'', where this project was initiated.

\appendix

\section{Other wash-out processes}
\label{App:scatterings}
It is known that for thermal leptogenesis, $\Delta L=1$ interactions $\gamma(L N \to Q_3t)$, $\gamma(t L \to N Q_3)$ as well as  $\Delta L=2$ interactions $\gamma(H^c L \to \bar L H)$ can bring numerical modifications to the final lepton asymmetry. 
In this Appendix, we would like to show that off-shell $2 \leftrightarrow 2$ scatterings are subdominant and can be safely discarded. 

The process $\gamma(H^c L \to \bar L H)$ decouples at temperatures below $T \lesssim 10^{13}\,\GeV$ so will we ignore it. The box-diagram-induced process, $\gamma(t L \to N Q_3)$, can be written as\,\cite{Davidson:2008bu}
\bea
\gamma (tL \to N Q_3) = \frac{g_t g_L T }{32 \pi^4} \int_{M_N^2} ds s^{3/2} K_1 (\sqrt{s}/T) \sigma_{tL \to N Q_3}(s).
\eea 
For $\sigma = y_t^2 Y_D^2/8\pi s$ and defining $z \equiv \frac{M_N}{T} \gg 1$, we obtain 
\begin{align}
\gamma (tL \to N Q_3) =& \frac{g_t g_L T }{32 \pi^4 }  \frac{y_t^2 Y_D^2}{8\pi} \int_{M_N^2} ds s^{1/2} K_1 (\sqrt{s}/T) 
\\
\approx &  \frac{g_t g_L T }{32 \pi^4} \frac{y_t^2 Y_D^2}{8\pi}\int_{M_N^2} ds \sqrt{\frac{\pi  \sqrt{s}}{2}} e^{-\sqrt{s}/T} 
\\
\approx & \frac{g_t g_L y_t^2 Y_D^2 }{128 \pi^5 } T^4 \sqrt{z}K_1(z).
\label{eq:rate2to2}
\end{align}
Comparing Eq\eqref{eq:rate2to2} with Eq\eqref{eq:rate_ID} we see that 
\bea 
\frac{\gamma (tL \to N Q_3)}{\gamma (HL \to N )} \approx  0.1 z^{-3/2} . 
\eea 
We then conclude that we can neglect the $(tL \to N Q_3)$ in the Boltzmann equations. More generally, we can neglect Higgs-mediated off-shell $2 \to 2$ processes, because the Higgs propagator scales like $1/M_N^2$ in this regime and is then subdominant with respect to decays.  
Since the scalar field $\phi$ is light (and possibly the gauge field $A_\mu$), it is also abundant in the plasma even after the transition. As a consequence, another possible wash-out is $\gamma(\phi\phi \to NN)$, ($\gamma(A_\mu A_\mu \to NN)$), which has the rate
    \begin{align}
    \gamma_{\phi\phi \to NN} \approx & \frac{ T}{32\pi^4} \int_{s = (2M_N)^2}^\infty ds s^{3/2} K_1(\sqrt{s}/T) \sigma_{\phi\phi \to NN}
    \\
    \approx & \frac{\lambda_\phi^2 y^2 T}{128 \pi^5 \times 4} \int_{s = (2M_N)^2} ds s\sqrt{\pi \sqrt{s}/2} T e^{- \sqrt{s}/T}
    \\
    \approx & \frac{\lambda_\phi^2 y^2  T^4}{128 \pi^5 \times 4 }  \Gamma[4, 2z]. 
    \end{align} 
    where $\lambda_\phi$ is a loop induced quartic. Similarly 
    \bea
    \gamma_{AA \to NN} \approx \frac{g^2 y^2 T^4}{128 \pi^5 \times 4 }  \Gamma[4, 2z].
    \eea
    Requiring that this rate is smaller than the inverse decay, we obtain 
    \bea 
    \gamma_{ID} > \gamma_{\phi\phi \to NN}, \gamma_{AA \to NN}  ~ \Rightarrow ~  (\lambda_\phi^{-2}, g^{-2}) \gtrsim  \frac{1}{10^3 Y_D^2}\frac{\Gamma[4, 2z]e^{-z}}{z^{5/2}}
    \label{eq:dec_phiphi}
    \eea

\section{Decay of the light $\phi$}
\label{App:phi_decay}
In all the parameter space we study, the scalar field is a light dof with loop-suppressed mass 
\bea 
m_\phi^2 &=& \beta_\lambda e^{-(2\delta \lambda/\beta_\lambda+1/2)} \mu_*^2
= \beta_\lambda v_\phi^2
\ll m_N^2, m_s^2, m_A^2.
\eea 
The channel of decay $\phi \to NN, ss$, or $AA$ is thus kinematically forbidden. 

However, $\phi$ cannot be stable because we cannot forbid its Higgs portal interaction,
\bea
-\Delta {\cal L} = \lambda_{h\phi} |H|^2 |\Phi|^2.
\label{eq:Higgsportal}
\eea
Since it recieves quantum corrections from the box diagram of $L$ and $N$,
even if we assume $\lambda_{h\phi}=0$ at some RG scale, it becomes nonzero at other scales, so $|\lambda_{h\phi}| \gsim y^2 Y_D^2/16\pi^2$. 
However, this lower bound of $\lambda_{h\phi}$ implies that there must be a fine tuning from the bare Higgs mass term since Eq.\,\eqref{eq:Higgsportal} gives a large contribution to the Higgs quadratic term with $\Delta m_h^2 = \lambda_{h \phi} v_\phi^2 $.
Therefore, we need a large tuning, anyway, so we give up setting $\lambda_{h\phi} $ to be small.

Taking $\lambda_{h\phi}$ to be a free parameter, the decay rate of $\phi \to H^\dagger H$ can be obtained as
\bea
\Gamma(\phi \to H^\dagger H) \sim \frac{|\lambda_{h\phi}v_\phi|^2}{8\pi \, m_\phi}.
\eea
Comparing this to the Hubble rate, we obtain
\bea
\frac{\Gamma}{H} \sim 
\mathcal{O}(1) 
\left( \frac{\lambda_{h\phi}^2}{10^{-4}} \right)^{2}
\left( \frac{10^{-2}}{\beta_{\lambda}} \right)^{1/2} 
\left( \frac{v_\phi}{10^9\,\GeV} \right)
\left( \frac{10^9\,\GeV}{T} \right)^2,
\eea
so $\phi$ decays rapidly if $\lambda_{h\phi}>10^{-4}$, which is small enough not to mess up the phase transition properties we obtained in the main text.

\section{Gravitational wave signal}
\label{App_GW}
In this appendix, we will quickly review the expressions for the gravitational signal induced by the phase transition. Theoretically, two different sources of GW are well understood; the \emph{bubble collision}\cite{Cutting:2018tjt}, dominating the signal in the case of runaway walls (theories with no gauge bosons)\cite{Bodeker:2009qy}, and the \emph{plasma sound wave}\cite{Caprini:2019egz}, dominating in the case of terminal velocity walls, (theories with gauge bosons)\cite{Bodeker:2017cim}.

\subsection{Energy budget }

As the transition completes, it releases energy, which can go into sound waves propagating in the plasma, heat, and kinetic energy accelerating the bubble walls. To manifest the conservation of energy, we define the following parameters:
 \begin{equation}
 \kappa_{\text{wall}} ,\quad\mbox{and}\quad \kappa_{\text{fluid}} = 1 - \kappa_{\text{wall}}. 
 \label{parameters}
 \end{equation}
The parameter $\kappa_{\text{wall}}$ can be understood as a measure of the ratio of energy going to the wall kinetic energy
\begin{equation}
\kappa_{\text{wall}} \equiv \frac{E_{\text{wall}}}{E_{\text{total}}}
\end{equation}
and depends on the regime of the velocity of the bubble wall. As we have seen above, the regime of expansion of the bubble results from the balance between the driving force $\Delta V$ and the pressure originating from the plasma
\bea
\mathcal{P}_{\rm tot} \approx \underbrace{\mathcal{P}_{\rm LO}}_{\eqref{eq:BPpres}} + \underbrace{\mathcal{P}_{\text{NLO}}}_{\eqref{NLO}}. 
\eea 
where the first contribution in the LO (leading order) contribution to the pressure and the second term is the NLO (next-to-leading order) contribution to the pressure. 
The condition 
\bea 
\Delta V = \mathcal{P}_{\rm tot}(\gamma= \gamma_{\rm terminal })
\label{eq:terminal}
\eea 
defines the terminal boost factor of the bubble. 
On the other hand, if Eq.~\eqref{eq:terminal} is never fulfilled, bubble walls get accelerated until the percolation, and we can estimate the boost factor at collision as
\begin{equation}
	\gamma_{\star} \sim \frac{R_\star}{R_c}
\end{equation}
with $R_c = \Big(\frac{3}{2\pi}\frac{S_3}{\Delta V}\Big)^{1/3} \sim 1/T_{\rm nuc}$ and $R_\star$ the size of the bubble at the collision. 	
In our work, there are two qualitatively different regimes as follows\,\footnote{
Remind that we take the benchmark parameters in the study of gravitational waves to have the enhancement of baryon asymmetry maximized for a given $y$.
Consequently, all the benchmark points have $\Delta V$ always greater than ${\cal P}_{\rm LO}$, and therefore, in any case, we approximate $v_w \simeq 1$.
}.

\begin{enumerate}
	\item {\bf Relativistic but with a terminal velocity (GBC)} 
	\newline
	The driving force is large enough to overcome the leading order friction but not the NLO friction. As a sizeable $\gamma$ is reached, the NLO order contribution to the friction balances the latent heat and the acceleration of the wall ends, defining a terminal value $\gamma_{\text{terminal}}$. The two conditions for this regime take the form:
\bea 
 \Delta V > \mathcal{P}_{\text{LO}}, \quad\mbox{and}
 \quad  \Delta V =  \mathcal{P}_{\text{NLO}}\big|_{\gamma = \gamma_{\text{terminal}}}. 
\eea
In this context, the parameters defined above become
\begin{equation}
 \kappa_{\text{wall}} = 0 , \quad\mbox{and}\quad
 \kappa_{\text{fluid}} = 1 . 
 \label{parameters3}
 \end{equation}

 \item {\bf Runaway Regime (SC)} 
\newline
When the symmetry involved in the PT is not gauged, $\Delta m_{\rm GB}=0$ in Eq.\,\eqref{NLO}.
Then, the release of energy $\Delta V$ is large enough to overcome all the sources of friction and then the wall keeps accelerating until the collision. 
Mathematically, the condition can be written as
\bea 
 \Delta V > \big(\mathcal{P}_{\text{LO}}
 +\mathcal{P}_{\text{NLO}}\big) |_{\rm collision}
\eea 
In this case, the energy budget parameters introduced above become
\begin{equation}
 \kappa_{\text{wall}} = 1 - \frac{\alpha_{\infty}}{\alpha_n}, \quad\mbox{and}\quad \kappa_{\text{fluid}} = \big(1 - \kappa_{\text{wall}}\big)
 \label{parameters2}
 \end{equation}
 where 
 \bea
 \alpha_{\infty} = \frac{\mathcal{P}_{\text{LO}}}{\rho_{\text{radiation}}}. 
 \eea
\end{enumerate}

\subsection{From bubble collision}

The amplitude and spectrum of the GW signal from bubble collision has been simulated in\cite{Cutting:2018tjt}
\bea
\frac{d\Omega_{\phi}h^2}{d\text{ln}(f)} 
&=& 
    3.22\times 10^{-3} F_{{\rm gw},0}h^2 (H_{\rm reh}R_\star)^2\bigg(\frac{\kappa_{\text{wall}}\alpha_n }{1+\alpha_n}\bigg)^2S_{\phi}(f, \tilde{f}_\phi),
\eea
where $F_{{\rm gw},0}\simeq 3.5\times 10^{-5}\,(100/g_*)^{1/3} $ is the red-shift factor of the radiation until today with $h=0.67$\,\cite{Planck:2018vyg},
$g_\star$ is the number of relativistic degrees of freedom at $T_{\rm reh}$,
$H_{\text{reh}}=\sqrt{\frac{8\pi}{3} \frac{\pi^2}{30}g_\star}T_{\rm reh}^2/M_{\rm Pl}$ is the Hubble rate evaluated at $T_{\rm reh}$ and $R_\star$ is the average radius of the bubbles at the percolation.
The spectral function $S_\phi$ is given by 
\begin{equation}
S_{\phi}(f,\tilde{f}) = \frac{(a+b)^c \tilde{f}^bf^a}{\big(b\tilde{f}^{\frac{a+b}{c}}+a f^{\frac{a+b}{c}}\big)^c}
\end{equation}
with $ a =3,~  b = 1.51,~ c = 2.18$, and the peak frequency 
\bea
\tilde{f}_\phi &=& 
\frac{3.2}{2\pi R_\star}
\times
\l(
    3.7\times 10^{-5}
    \, g_\star^{-1/3}
    \l( \frac{100\,\GeV}{T_{\rm reh}} \r)
\r)
\\
&=&  0.35 \times 10^{-5}
\,
\beta_{\rm PT} \,
\bigg(\frac{T_{\rm reh}}{100\,\GeV} \bigg)\bigg( \frac{g_\star}{100}\bigg)^{1/6}
\text{ Hz},
\eea
where we approximate
\bea
R_\star \simeq \frac{(8\pi)^{1/3}}{\beta_{\rm PT} H_{\rm reh}}.
\eea
Remind that $\beta_{\rm PT}$ is dimensionless in our definition.

\subsection{From sound waves}
The gravitational wave signal induced by sound waves is given by\cite{Hindmarsh:2017gnf,Ellis:2018mja}
\bea
\frac{d\Omega_{\rm sw}h^2}{d\ln(f)}
&=&
2.061 \,
F_{\rm gw,0}h^2 \,
\Gamma^2 \,
\bar U_f^4 \,
(H_n R_\star) \,
\tilde \Omega_{\rm gw}\,
S_{\rm sw}(f/f_{p,0})
\times
\min\left(1,H_{\rm reh}R_\star/\bar U_f
\right)
\label{eq:GW_sw}
\eea
where $\Gamma=1+\bar p/\bar \epsilon \simeq 4/3$ is the adiabatic index, 
$\bar U_f$ is the RMS fluid velocity\,\cite{Hindmarsh:2015qta},
$\tilde \Omega_{\rm gw}$ is a dimensionless efficiency factor.
The last term of Eq.\,\eqref{eq:GW_sw}
takes into account the suppression factor in the regime of $H_{\rm reh}R_\star/\bar U_f<1$
as discussed in Ref.\,\cite{Ellis:2018mja}.
We approximate $\bar U_f$ by following Ref.\,\cite{Hindmarsh:2015qta}
\bea
\bar U_f \simeq \sqrt{\frac{3}{4}\frac{\kappa_{\rm sw}\alpha_n}{1+\alpha_n}},
\eea
where\,\cite{Hindmarsh:2015qta,Espinosa:2010hh}
\bea 
\kappa_{\rm sw} \simeq
\kappa_{\rm fluid}\frac{\alpha_n}{0.73 + 0.083 \sqrt{\alpha_n}+ \alpha_n}.
\eea 
We take $\tilde \Omega_{\rm gw} \sim 10^{-2}$ from the numerical simulation\,\cite{Hindmarsh:2017gnf, Cutting:2019zws} in the case of detonations, and depict the result in Fig.\,\ref{fig:GW} by solid and dashed lines, respectively.
The peak frequency is given by
\bea
\tilde f_{\rm sw}
=2.6\times 10^{-5}
\l(\frac{1}{H_{\text{reh}} R_\star}\r)
\l(\frac{z_p}{10}\r)
\l(\frac{T_{\rm reh}}{100 \text{ GeV}}\r)
\l(\frac{g_\star}{100}\r)^{1/6} \text{ Hz} ,
\eea
where $z_p$ parametrizes the actual peak position in the numerical simulation.
We take $z_p\simeq 10$
based on Ref.\,\cite{Hindmarsh:2017gnf}.
The spectral shape of GW is given by
\bea
 S_{\rm sw} = s^3 \bigg(\frac{7}{4+3s^2}\bigg)^{7/2}.
 \label{eq:spectrum}
\eea
\\
Since the numerical simulations of gravitational waves from the sound waves have been only performed up to $\alpha_n \lsim 0.3$, we set a relatively large uncertainty range of $\tilde \Omega_{\rm gw}$ in Eq.\,\eqref{eq:GW_sw} since we focus on $\alpha_n \sim 5$.
We find that, in terms of the amplitude at the peak frequency, the sound wave contribution is more dominant compared to the scalar contribution within the uncertainty of $\tilde \Omega_{\rm gw}$, which can also be found in other literature\,\cite{Ellis:2020nnr}.

\subsection{Comments on other sources}
During a FOPT, there are other sources that can provide additional gravitational wave signals.
For instance, turbulences made from bubble collisions can provide additional gravitational waves\,\cite{Gogoberidze:2007an}, but we do not take it into account because the current uncertainty in the numerical studies is large.

Another interesting idea was recently proposed in Ref.\,\cite{Jinno:2022fom} where the particles inside bubbles freely stream within the time scale of the PT duration.
In this case, the peak frequency can be shifted to a lower value compared to the sound wave contribution.
Although this effect might increase the testability of our scenario, for $M_N \gsim 10^8\,\GeV$, the decay of $N_I$ dominates, and the interactions of their daughter particles do not seem sufficiently feeble (because they are SM particles).
On the other hand, when $M_N\lsim 10^8\,\GeV$, we have a large depletion from the annihilation $N_I N_I \to \phi\phi$, so the most of the energy density can be transferred to $\phi$ which is long-lived and feebly interacting with the plasma at $T_n \ll M_N$.
In this case, we may be able to apply the scheme of Ref.\,\cite{Jinno:2022fom}, but this parameter space is less motivated in the explanation of the observed baryon asymmetry.

\bibliographystyle{JHEP}
{\footnotesize
\bibliography{biblio}}
\end{document}